\documentclass[%
 reprint,
 amsmath,amssymb,
 aps,
]{revtex4-2}

\usepackage{graphicx}% Include figure files
\usepackage{dcolumn}% Align table columns on decimal point
\usepackage{bm}
\usepackage{mathrsfs}
\usepackage{amsmath}
\usepackage{enumitem} 
\usepackage{cancel,siunitx}
\let\svqty\qty
\usepackage{physics}
\let\qty\svqty
\usepackage[]{xcolor}
\usepackage{soul}

\begin{document}

\preprint{APS/123-QED}

\title{Strong coupling in chiral cavities: \\
nonperturbative framework for enantiomer discrimination}% Force line breaks with \\
%\title{Modeling strong coupling in chiral cavities}% Force line breaks with \\

\author{Rosario R. Riso$^{1}$} 
\author{Laura Grazioli$^{2}$}
\author{Enrico Ronca$^{3}$}
\author{Tommaso Giovannini$^{4}$}
\author{Henrik Koch$^{1,4}$}
\email{henrik.koch@sns.it}
\affiliation{$^{1}$Department of Chemistry, Norwegian University of Science and Technology, 7491 Trondheim, Norway}
\affiliation{$^{2}$Institut für Physikalische Chemie, Universität Mainz, D-55099 Mainz, Germany}
\affiliation{$^{3}$Dipartimento di Chimica, Biologia e Biotecnologie, Università degli Studi di Perugia,
Via Elce di Sotto, 8,06123, Perugia, Italy}
\affiliation{$^{4}$Scuola Normale Superiore, Piazza dei Cavalieri 7, 56126 Pisa, Italy}

\date{\today}% It is always \today, today,

\begin{abstract}
Development of efficient techniques to distinguish mirror images of chiral molecules (enantiomers) is very important in both chemistry and physics. Enantiomers share most molecular properties except, for instance, the absorption of circularly polarized light. Enantiomer purification is therefore a challenging task that requires specialized equipment.
Strong coupling between quantized fields and matter (e.g. in optical cavities) is a promising technique to modify molecular processes in a non-invasive way. The modulation of molecular properties is achieved by changing the field characteristics. In this work, we investigate whether strong coupling to circularly polarized electromagnetic fields is a viable way to discriminate chiral molecules. To this end, we develop a nonperturbative framework to calculate the behaviour of molecules in chiral cavities. We show that in this setting the enantiomers have different energies -- that is, one being more stable than the other. The field-induced energy differences are also shown to give rise to enantiospecific signatures in rotational spectra.
\end{abstract}

\maketitle

%\tableofcontents
\section{Introduction}
The study of strongly coupled light-matter systems is becoming a well established area in both physics and chemistry\cite{flick2018strong,li2021molecular,ebbesen2016hybrid,garcia2021manipulating,hertzog2019strong,bloch2022strongly}. Through the interaction with the quantized field, it is indeed possible to modify both macroscopic material features, for example conductivity or phase transitions \cite{sentef2018cavity,ashida2020quantum}, or to influence microscopic properties like the chemistry of single molecules \cite{mandal2019investigating,fregoni2018manipulating,fregoni2020photochemistry}. 
The easiest way to reach the strong coupling regime is through optical cavities, composed of two highly reflective mirrors placed in front of each other \cite{loudon2000quantum}. 
Because of the mirrors, the electromagnetic fields inside the cavity must fulfill specific boundary conditions. The modulation of the boundary conditions allows for a fine tuning of the main field properties. For example, choosing a certain geometry of the device, specific sets of frequencies can be selected. On the other hand, an opportune engineering of the mirrors can also lead to changes in the shape of the field or selection of its polarization  \cite{viviescas2003field,plum2015chiral,favero2009optomechanics}. The confinement of photons in a small quantization volume leads to an increase in the light-matter interaction \cite{chikkaraddy2016single,fitzgerald2016quantum} with the consequent formation of hybrid states called polaritons \cite{flick2017atoms}. Polaritons represent an effective way to modulate matter properties in a non-invasive way. In fact, the characteristics of the mixed electron-photon states can be manipulated by tuning the field properties \cite{sidler2020polaritonic,galego2015cavity}. In recent years, experimental developments have dramatically improved our control of the cavity field, opening the way to a wide range of new applications \cite{plum2015chiral,chikkaraddy2016single,viviescas2003field,forbes2021orbital,galego2019cavity, hutchison2012modifying, haugland2021intermolecular, riso2022characteristic,vergauwe2016quantum,lather2019cavity,thomas2019tilting,thomas2016ground}. Theoretical modeling of strong coupling phenomena is urgently needed to advance our intuitive understanding of what happens inside cavities and to assist the experiment-design phase. Many remarkable efforts have already been presented in the literature, with the introduction of both \textit{ab initio} electronic structure methods \cite{schafer2021making,haugland2020coupled,riso2022molecular,mandal2020polarized,ashida2021cavity,deprince2021cavity,galego2019cavity} and molecular dynamics techniques \cite{flick2020ab,schafer2021shining} for strongly coupled systems. Nonetheless, a comprehensive framework for arbitrary field shapes is still not available and requires further theoretical developments. 

One of the most intriguing perspectives in molecular polaritonics is the possibility of enhancing the spectroscopic techniques through the interaction with quantized fields. Pioneering works have indeed shown that upon coupling with the quantized fields it is possible to increase the spectral resolution by amplifying the signal intensities \cite{petryayeva2011localized,maccaferri2021recent}, even at the level of single molecule imaging \cite{yang2020sub, romanelli2021role, zhang2013chemical}. In this regard, a particularly interesting perspective is to use strong coupling to discriminate among chiral molecules (systems that are non-superimposable with their mirror image), for example through the formation of chiral polaritonic states \cite{li2022strong,mauro2022chiral,schafer2022chiral,sun2022polariton,gautier2022planar}. The two different configurations of a chiral molecule, called enantiomers, share most physical properties and can be distinguished when they interact with circularly polarized light \cite{schnoering2021chiral,sato2004asymmetric}. In particular, chiral molecules interact and absorb differently left-handed circular polarized (LHCP) and right-handed circular polarized (RHCP) light \cite{barron2004molecular,craig1998molecular}. This suggests that in chiral cavities  \cite{hubener2021engineering,plum2015chiral,schafer2022chiral,baranov2022towards,mauro2022chiral,li2022strong}, devices where the electromagnetic field has a fixed circular polarization, it might be possible to create energy differences between two enantiomers even in the ground state.\\

In this work, we present an \textit{ab initio} theoretical framework to study the strong coupling regime in chiral cavities. We first give a general introduction to chiral cavities and discuss the formal properties that a cavity quantum electrodynamical (cQED) model must obey. Afterwards, we develop a novel \textit{ab initio} methodology that enables the study of such systems. We demonstrate that, even when no real photons are present in the cavity, coupling to a circularly polarized field induces energy differences between the enantiomers. Finally, we show that the energy differences produce enantiospecific signatures in the molecular rotational spectra.

\section{Chiral cavities}
\begin{figure}
    \centering
    \includegraphics[width=0.45\textwidth]{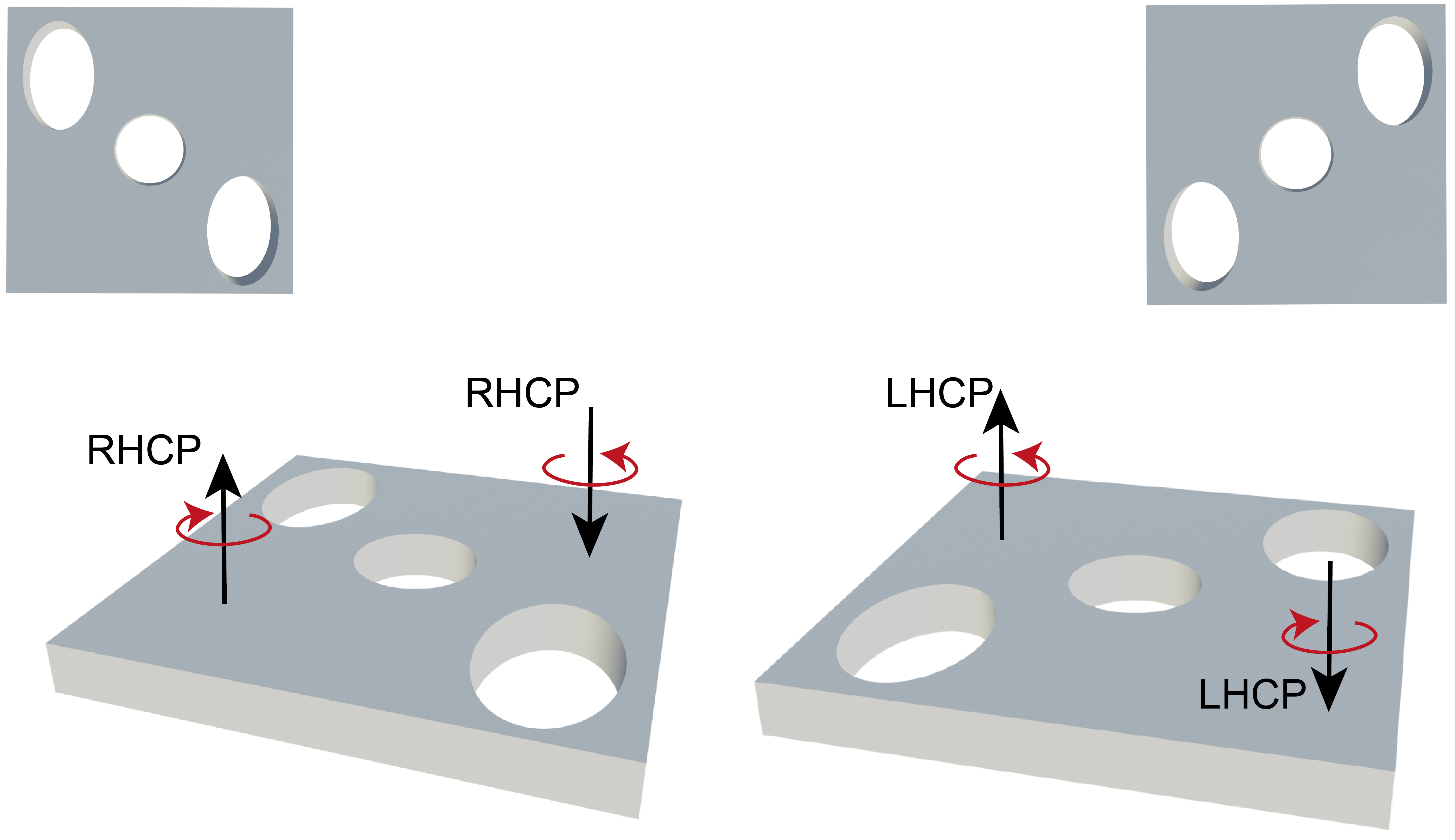}
    \caption{Example of a chiral pattern as seen from the two opposite sides of the metasurface. No in plane rotation can superimpose the two pattern. }
    \label{fig:Pattern}
\end{figure}

Chiral cavities are optical devices that allow only one circular polarization of light to exist within a certain volume (quantization volume). Construction of these structures has been a significant challenge for researchers over the past 10 years \cite{genet2022chiral,plum2015chiral,fedotov2006mirror}. Indeed, the mirrors of a chiral cavity must fully reflect light of one circular polarization and preserve the field handedness, the property that describes the direction of rotation of the electric field vector. At the same time, they must absorb or transmit the fields of the opposite circular polarization \cite{plum2015chiral,sun2022polariton,voronin2022single,mauro2022charge} 
%\revER{(The previous sentence is very long I would break it after field handedness!!)}.
These are quite stringent constraints. The most used metallic mirrors, for example, can not differentiate among the fields handedness \cite{plum2015chiral,fedotov2006mirror}. Moreover, the circular polarization of the wave is reversed upon reflection. Indeed, normal mirrors usually reverse the direction of either the electric, $\mathbf{E}$, or the magnetic, $\mathbf{B}$, field. Since the angular momentum of light, $\mathbf{J}$, is equal to \cite{jackson1999classical}
\begin{equation}
\mathbf{J}=\frac{1}{4\pi}\int \mathbf{r}\times \left(\mathbf{E}\times\mathbf{B}\right)d^{3}r, 
\end{equation}
the handedness is reversed if either  $\mathbf{E}$ or $\mathbf{B}$ changes sign. %This implies that not even shining circularly polarized light inside a Fabry-Pèrot cavity - the most common type of optical resonator - addresses the circular polarization problem \cite{schafer2022chiral,taradin2021chiral}. Handedness preservation upon reflection has been achieved for the first time in 2006 by Fedotov et al. using anisotropic metamaterial surfaces which allow for a $\pi$ shift in the phase of both the electric and magnetic field \cite{fedotov2006mirror}. 
%The main idea in the realization of a handedness preserving mirror is to place a metamaterial with some suitable properties in front of a conventional mirror. Chiral mirrors have been constructed following the same scheme.
The main idea in the realization of a chiral mirror is to place a metamaterial with some convenient properties in front of a conventional mirror. Differentiation between LHCP and RHCP light is achieved using optically active media. Three dimensional optically active media are not suitable for chiral mirrors \cite{plum2015chiral,gautier2022planar,semnani2020spin,voronin2022single}. 
%as the handedness reversal of the wave upon reflection through the metal mirror would erase any circular polarization discrimination obtained passing through the chiral material the first time.
The desired additional layer, instead, is a 2 dimensional chiral metamaterial that inverts its transmission and reflection properties based on the propagation direction of the wave. %Two dimensional chiral metamaterials
%, obtained using a pattern that can not be superimposed with its mirror image without being lifted from the plane, 
%posses the required property \revER{(This sentence is very complicated)}. 
These objects are obtained using patterns that can not be superimposed with their mirror images without being lifted from the plane. Since the pattern inverts its chirality on the two opposite sides of the surface, such structures preferentially reflect, for example, LHCP light on one side and RHCP light on the other, increasing the handedness selectivity of the chiral mirror. An example of such a pattern, taken from an experimentally realized mirror \cite{taradin2021chiral,semnani2020spin}, is shown in Figure \ref{fig:Pattern}.
A detailed and exhaustive discussion of the response properties of such mirrors can be found in Refs.\cite{mauro2022charge,mauro2022chiral}. Placing two chiral mirrors one in front of the other allows for the creation of chiral cavities. While so far many works have mainly focused on the creation of increasingly efficients chiral mirrors, experimental demonstrations of chiral cavities have also been presented in recent years, i.e. from Voronin et al. \cite{voronin2022single} and from Tarandin et al. \cite{taradin2021chiral}. Alternative possibilities for the production of chiral cavities that do not use chiral mirrors, have been recently proposed by Gautier et al. \cite{gautier2022planar}. In particular, their optical device is composed of a normal set of Fabry-Pérot mirrors with a layer of the 2D chiral polystyrene inserted at the middle of the cavity. Using the unique properties of the 2D chiral objects, the authors manage to isolate two regions of space where only one field handedness is present. \\

\section{Theoretical modeling}\label{sec:Theoretical}
The interaction between photons and matter will be described using the minimal coupling Hamiltonian \cite{wang2019cavity,schafer2018ab}
\begin{equation}
\begin{split}
H =&\frac{1}{2}\sum_{i}\left(\mathbf{p}_{i}-\mathbf{A}(\mathbf{r}_{i})\right)^{2}+ \sum_{I}\frac{1}{2M_{I}}\left(\mathbf{p}_{I}+Z_{I}\mathbf{A}(\mathbf{R}_{I})\right)^{2}\\
+&\sum_{i>j}\frac{1}{\left|r_{i}-r_{j}\right|}+\sum_{I>J}\frac{Z_{I}Z_{J}}{\left|R_{I}-R_{J}\right|}-\sum_{i, I}\frac{Z_{I}}{\left|R_{I}-r_{i}\right|}  \\
+&\frac{1}{2}\int\left(\mathbf{E}^{2}(\mathbf{r})+c^{2}\mathbf{B}^{2}(\mathbf{r})\right)d^{3}r,
\label{eq:Minimal}
\end{split}
\end{equation}
where $i$ and $j$ label electrons while $I$ and $J$ label nuclei with charges $Z_{I}$ and $Z_{J}$. The vector potential, the electric and the magnetic fields are denoted by $\mathbf{A}(\mathbf{r})$, $\mathbf{E}(\mathbf{r})$ and $\mathbf{B}(\mathbf{r})$, respectively. Since in the strong coupling regime the electromagnetic field is a critical component of the system, it must be treated on the same footing as the electrons by means of quantum electrodynamics (QED) \cite{craig1998molecular}. The optical properties of a cavity are encoded in the vector potential $\mathbf{A}(\mathbf{r})$ (and consequently in $\mathbf{E}(\mathbf{r})$ and $\mathbf{B}(\mathbf{r})$).
Specifically, in a chiral cavity only one of the two possible circular polarizations of the field is allowed. The second quantized vector potential can therefore be written as
\begin{equation}
\mathbf{A}(\mathbf{r}) =\sum_{k} \frac{\lambda}{\sqrt{2\omega_{k}}}\left(\boldsymbol{\epsilon}_{k\pm}b_{k}e^{i\mathbf{k}\mathbf{r}}+\boldsymbol{\epsilon}^{*}_{k\pm}b^{\dagger}_{k}e^{-i\mathbf{k}\mathbf{r}}\right),  
\label{eq:Shape}
\end{equation}
where $k$ spans all the possible wave vectors and $\boldsymbol{\epsilon}_{k\pm}$ denotes the field polarization. In Eq.~(\ref{eq:Shape}), $\omega_{k}$ is the frequency of the field. The parameter $\lambda$ quantifies the strength of the light-matter coupling. This quantity is related to the quantization volume of the electromagnetic field, V. In particular $\lambda = \sqrt{\frac{\hbar}{\epsilon_{0}V}}$. 
The coupling value significantly affects the magnitude of the cavity induced effects, i.e. large $\lambda$ imply strong field effects. When $\lambda$ is zero, instead, the cavity and the molecule are completely decoupled. For this reason, a lot of research is devoted towards the confinement of electromagnetic fields in smaller volumes. New experimental techniques keep pushing the boundaries of the ultrastrong coupling regime by extreme confinement of the field quantization volume, using for instance nanoplasmonic picocavities \cite{carnegie2018room, baumberg2022picocavities}. In some cases, the cavity volume, and therefore the quantization volume, can be reduced below to the $nm^{3}$ limit, i.e. $\lambda>0.05\;au$ \cite{chikkaraddy2016single}. Reported realizations of chiral cavities have so far only reached couplings of $0.01 \;au$ \cite{sun2022polariton,voronin2022single}. However, there are no theoretical limitations that prevent us from reducing the quantization volume of such fields below the $nm^{3}$ limit. In this paper we will assume, without loss of generality, that wave vector $\mathbf{k}$ is aligned along the z-axis and the field polarization is  
\begin{equation}
\boldsymbol{\epsilon}_{k\pm}=\boldsymbol{\epsilon}_{\pm}=\frac{1}{\sqrt{2}}\begin{pmatrix}
1 \\ \pm i \\ 0
\end{pmatrix}.    
\end{equation}
In the Born-Oppenheimer approximation, Eq.~(\ref{eq:Minimal}) takes the form 
\begin{equation}
\begin{split}
H =& \frac{1}{2}\sum_{i}\left(\mathbf{p}_{i}-\sum_{k}\frac{\lambda}{\sqrt{2\omega_{k}}}\left(\boldsymbol{\epsilon}_{\pm}b_{k}e^{i\mathbf{k}\mathbf{r}_{i}}+\boldsymbol{\epsilon}^{*}_{\pm}b^{\dagger}_{k}e^{-i\mathbf{k}\mathbf{r}_{i}}\right)\right)^{2}\\
+&\sum_{i>j}\frac{1}{\left|r_{i}-r_{j}\right|}+\sum_{I>J}\frac{Z_{I}Z_{J}}{\left|R_{I}-R_{J}\right|}-\sum_{i, I}\frac{Z_{I}}{\left|R_{I}-r_{i}\right|}\\
+&\sum_{k}\omega_{k}\left(b^{\dagger}_{k}b_{k}+\frac{1}{2}\right),  
\label{eq:Multimode}
\end{split}
\end{equation}
where we have used that the nuclear mass $M_{I}>c$ in atomic units and the nuclei are kept fixed. 
Since the full Hamiltonian in Eq.~(\ref{eq:Multimode}) involves an infinite number of modes, it is computationally unfeasible to determine the eigenvalues and eigenfunctions. However, we expect that restricting Eq.~(\ref{eq:Multimode}) to include only one field frequency, i.e. two modes, will be sufficient to obtain a qualitatively correct description of the field effects. 
Upon truncation, the Hamiltonian for a molecular system in the chiral cavity (Eq.~(\ref{eq:Multimode})) is
\begin{equation}
\begin{split}
H =& \sum_{i}\frac{\mathbf{p}^{2}_{i}}{2}+\frac{N_{e}\lambda^{2}}{2\omega}\left(b_{k}+b^{\dagger}_{-k}\right)\left(b_{-k}+b^{\dagger}_{k}\right)\\
+&\sum_{i>j}\frac{1}{\left|r_{i}-r_{j}\right|}+\sum_{I>J}\frac{Z_{I}Z_{J}}{\left|R_{I}-R_{J}\right|}-\sum_{i, I}\frac{Z_{I}}{\left|r_{i}-R_{I}\right|}\\
+&\frac{\lambda}{\sqrt{2\omega}}\sum_{i}\left(\mathbf{p}_{i}\cdot\boldsymbol{\epsilon}_{\pm}\right)(b_{k}+b^{\dagger}_{-k})e^{i\mathbf{k}\mathbf{r}_{i}}\\
+&\frac{\lambda}{\sqrt{2\omega}}\sum_{i}(\mathbf{p}_{i}\cdot\boldsymbol{\epsilon^{*}}_{\pm})(b^{\dagger}_{k}+b_{-k})e^{-i\mathbf{k}\mathbf{r}_{i}}\\
+&\omega\left(b^{\dagger}_{k}b_{k}+b^{\dagger}_{-k}b_{-k}+1\right),
\label{eq:Twomodes}
\end{split}
\end{equation}
where $N_{e}$ is the number of electrons in the system and and $\omega$ is used to denote $\omega_{k}$.
\begin{figure*}
    \centering
    \includegraphics[width=0.85\textwidth]{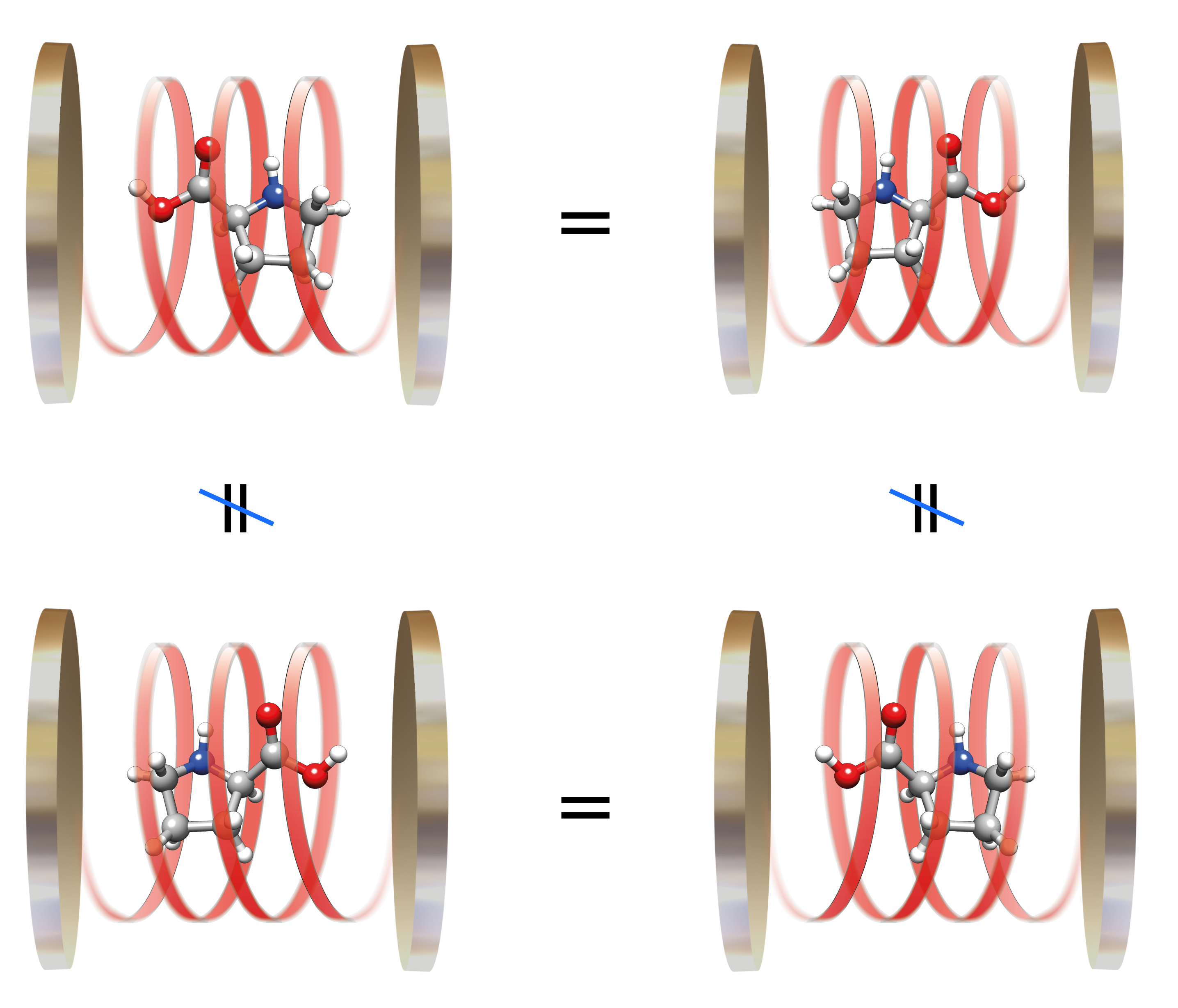}
    \caption{The L enantiomer of proline in a cavity with a LHCP field has the same energy as the R enantiomer in a cavity with a RHCP. However, no symmetry in the Hamiltonian ensures that the L enantiomer has the same energy in the LHCP cavity as in a RHCP polarized cavity.}
    \label{fig:RS_SR}
\end{figure*}
From the properties of the Hamiltonian in Eq.~(\ref{eq:Twomodes}), we can infer some of the characteristic features of a chiral cavity. Upon a reflection of the total system (molecule plus cavity) the molecule transforms into its mirror image while the cavity polarization is inverted ( LHCP$\leftrightarrow$RHCP).
The energy of the system remains the same because there is no parity-violating interaction in the Hamiltonian \cite{appelquist1986spontaneous}. However, while a non-chiral molecule is indistinguishable from its mirror image, performing a reflection of a chiral system will produce a different enantiomer, non-superimposable with the original molecule. Therefore non-chiral molecules have the same energy inside two differently circularly polarized cavities. On the other hand, a chiral molecule in a chiral cavity, e.g. a LHCP cavity, has the same energy as the other enantiomer in a cavity with the opposite polarization, e.g. a RHCP cavity. 
Most importantly, there is no symmetry of the Hamiltonian that requires the two enantiomers to have the same energy in the same chiral cavity, unlike in the vacuum case. Strong coupling between circularly polarized light and chiral molecules might therefore be a viable way to create a energy difference between enantiomers, as shown in Fig.~\ref{fig:RS_SR}. We refer to these cavity induced energy differences as the discrimination power of the cavity. \\
The discriminating properties are lost when the dipole approximation is adopted, even if the field polarization remains chiral. This can be verified by setting $e^{i\mathbf{k}\mathbf{r}}$ equal to one in Eq.~(\ref{eq:Twomodes}). In this case, the Hamiltonians for the LHCP and RHCP cavities have the same eigenvalues, as one Hamiltonian can be transformed into the other through a relabeling of the wave vectors ($\mathbf{k}\leftrightarrow-\mathbf{k}$). Thus, the exponential $e^{i\mathbf{k}\mathbf{r}}$ in the field parameterization plays a critical role for the discriminating power of the cavity. 
Modeling the chiral nature of the electromagnetic field, therefore, requires that we go beyond the dipole approximation in the description of the cavity field. These effects have not been included in \textit{ab initio} QED methodologies so far \cite{ruggenthaler2014quantum,mandal2020polarized,haugland2020coupled}. Moreover, due to the nature of the chiral field, it is essential to use a complex Hamiltonian. This is a delicate aspect to take into account when designing a new wave function approach as discussed in Sec.~\ref{sec:QED-CC}.\\
We will employ the Hamiltonian in Eq.~(\ref{eq:Twomodes}) to study a chiral molecule in a chiral cavity. The eigenvalue problem of this Hamiltonian becomes easier if we remove the quadratic term in the field. This can be accomplished by a Bogoljubov transformation \cite{rokaj2022free} that introduces two new bosonic operators $\alpha$ and $\beta$
\begin{equation}
\begin{split}
&\begin{pmatrix} b_{k} \\
b^{\dagger}_{-k} \\
\end{pmatrix} =
\begin{pmatrix} 
\cosh{\theta}  & -\sinh{\theta} \\
-\sinh{\theta} & \cosh{\theta} \\
\end{pmatrix}   \begin{pmatrix} \alpha \\
\beta^{\dagger} \\
\end{pmatrix}\\
&\begin{pmatrix} \alpha \\
\beta^{\dagger} \\
\end{pmatrix} =
\begin{pmatrix} 
\cosh{\theta}  & \sinh{\theta} \\
\sinh{\theta} & \cosh{\theta} \\
\end{pmatrix}   \begin{pmatrix} b_{k} \\
b^{\dagger}_{-k} \\
\end{pmatrix} ,
\label{eq:Transformation}
\end{split}
\end{equation}
that also satisfy the bosonic commutation relation
\begin{equation}
\begin{split}
[\alpha, \alpha^{\dagger}]=1 \hspace{1cm} [\alpha, \beta]=0 \hspace{1cm}
[\alpha, \beta^{\dagger}]=0.
\end{split}
\end{equation}
If $\tanh{2\theta}=\frac{N_{e}\lambda^{2}}{N_{e}\lambda^{2}+2\omega^{2}}$ in Eq.~(\ref{eq:Transformation}), the Hamiltonian can be rewritten as
\begin{equation}
\begin{split}
H =& \sum_{i}\frac{\mathbf{p}^{2}_{i}}{2}+\tilde{\omega}\left(\alpha^{\dagger}\alpha+\beta^{\dagger}\beta+1\right)\\
+&\sum_{i>j}\frac{1}{\left|r_{i}-r_{j}\right|}+\sum_{I>J}\frac{Z_{I}Z_{J}}{\left|R_{I}-R_{J}\right|}-\sum_{jI}\frac{Z_{I}}{\left|r_{j}-R_{I}\right|}\\   
+&\frac{\lambda}{\sqrt{2\tilde{\omega}}}\sum_{i}\left(\mathbf{p}_{i}\cdot\boldsymbol{\epsilon}_{\pm}\right)(\alpha+\beta^{\dagger})e^{i\mathbf{k}\mathbf{r}_{i}}\\
+&\frac{\lambda}{\sqrt{2\tilde{\omega}}}\sum_{i}(\mathbf{p}_{i}\cdot\boldsymbol{\epsilon}^{*}_{\pm})(\alpha^{\dagger}+\beta)e^{-i\mathbf{k}\mathbf{r}_{i}},
\label{eq:No_squared}
\end{split}
\end{equation}
where $\tilde{\omega}=\sqrt{\omega^{2}+N_{e}\lambda^{2}}$ and the $\mathbf{A}^{2}$ term in Eq.~(\ref{eq:Twomodes}) has been reabsorbed \cite{rokaj2022free}. Note that, since the transformation in Eq.~(\ref{eq:Transformation}) can be expressed as a unitary transformation of the photonic creation and annihilation operators (e.g. $\alpha=U^{\dagger}_{k}b_{k}U_{k}$) with the operator
\begin{equation}
U_{k} = \textrm{exp}\left[\theta\left(b^{\dagger}_{k}b^{\dagger}_{-k}-b_{k}b_{-k}\right)\right],   
\label{eq:Rotation}
\end{equation}
the Hamiltonians in Eqs.~(\ref{eq:Twomodes}) and (\ref{eq:No_squared}) have the same eigenvalues. The form in Eq.~(\ref{eq:No_squared}) is particularly interesting since the zero-point energy term explicitly depends on the number of electrons. This result is relevant for two main reasons:
\begin{enumerate}[label=(\roman*)]
    \item When the cavity frequency is approaching zero, numerical methods based on the minimal coupling approach in Eq.~(\ref{eq:Twomodes}) become unstable because the light-matter coupling term diverges as $\omega^{-\frac{1}{2}}$ while the quadratic term in the field diverges as $\omega^{-1}$. The Hamiltonian in Eq.~(\ref{eq:No_squared}) does not diverge for $\omega=0 \; \textrm{eV}$;
    \item The Hamiltonian in Eq.~(\ref{eq:No_squared}) is explicitly non size-extensive due to the contributions from $\tilde{\omega}$, which is non linear in the number of electrons. We note that in the limit of $\lambda=0\;au$ the Hamiltonian becomes size-extensivity as in the no cavity case.
\end{enumerate}
In Appendix A, we show that $\tanh{2\theta}=\frac{N_{e}\lambda^{2}}{N_{e}\lambda^{2}+2\omega^{2}}$ minimizes the zero-point energy and that the non size-extensive effects obtained from the zero-point energy contribution in Eq.~(\ref{eq:No_squared}) have the correct dependence on the number of electrons when $\lambda/\omega << 1$. 
The transformation in Eq.~(\ref{eq:Rotation}) is also useful when more modes are included in the minimal coupling Hamiltonian as shown in Appendix B.

\section{QED coupled cluster for the minimal coupling Hamiltonian}\label{sec:QED-CC}

We now present a QED coupled cluster (QED-CC) framework for the minimal coupling Hamiltonian \cite{liebenthal2022equation,haugland2020coupled,deprince2021cavity,pavosevic2021polaritonic,pavosevic2022cavity}. This approach is referred to as minimal coupling QED-CCSD (MC-QED-CCSD). The starting point is the electron-photon wave function
\begin{align}
\ket{\psi} =&\;\textrm{exp}\left((S_{1\alpha}+\gamma_{\alpha})\alpha^{\dagger}+(S_{1\beta}+\gamma_{\beta})\beta^{\dagger} \right)\nonumber\\
\times&\;\textrm{exp}\left(T_{1}+T_{2}\right)\ket{\textrm{HF}}\otimes\ket{0,0},  \label{eq:QED-CCSD-S-MINIMAL}
\end{align}
where $\ket{\textrm{HF}}$ is the no cavity Hartree-Fock Slater determinant while $\ket{0,0}$ denotes the photonic vacuum for the $\alpha$ and $\beta$ photons introduced in Eq.~(\ref{eq:Transformation}). 
The operators in the exponentials of Eq.~(\ref{eq:QED-CCSD-S-MINIMAL}) are electron ($T_{1}$ and $T_{2}$) and electron-photon ( $S^{\alpha}_{1}\alpha^{\dagger}$ and $S^{\beta}_{1}\beta^{\dagger}$) excitation operators defined explicitly as 
\begin{equation}
\begin{split}
&T_{1} = \sum_{ai}t^{a}_{i}E_{ai} \\
&T_{2}=\frac{1}{2}\sum_{abij} t^{ab}_{ij}E_{ai}E_{bj}\\
&S_{1\alpha}\alpha^{\dagger}=\sum_{ai}s^{a}_{i\alpha}E_{ai}\alpha^{\dagger} \\
&S_{1\beta}\beta^{\dagger}= \sum_{ai}s^{a}_{i\beta}E_{ai}\beta^{\dagger}.\label{eq:Excitation}
\end{split}
\end{equation}
The amplitude parameters $t^{a}_{i}$,$t^{ab}_{ij}$, $s^{a}_{i\alpha}$, $s^{a}_{i\beta}$ as well as $\gamma_{\alpha}$ and $\gamma_{\beta}$ are determined in the ground state calculation. 
In Eq.~(\ref{eq:Excitation}), we use second quantization for the electrons \cite{helgaker2014molecular} with
\begin{equation}
E_{pq} = \sum_{\sigma}a^{\dagger}_{p\sigma}a_{q\sigma},\label{eq:Second_quantized}    
\end{equation}
where $a^{\dagger}_{p\sigma}$ creates and $a_{p\sigma}$ annihilates an electron in orbital $p$ with spin $\sigma$.
The indices $i,j$ and $a,b$ label occupied and virtual orbitals in the HF reference, respectively.
In the limit where all excitations are included in the wave function in Eq.~(\ref{eq:QED-CCSD-S-MINIMAL}), the coupled cluster expansion is exact and gives the same result as QED full configuration interaction \cite{haugland2021intermolecular,haugland2020coupled,riso2022molecular}. The optimal values for the coupled cluster parameters are obtained by solving the projection equations \cite{helgaker2014molecular} 
\begin{equation}
\begin{split}
\Omega_{\mu,n,m}=&\bra{\mu,n,m}\textrm{exp}\left(-T\right)H\;\textrm{exp}\left(T\right)\ket{\textrm{HF},0,0}\\
=&0,
\label{eq:Omega}
\end{split}
\end{equation}
with
\begin{equation}
\begin{split}
&\ket{\textrm{HF},0,0}  = \ket{\textrm{HF}}\otimes\ket{0,0}\\[3pt]
&\ket{\mu,n,m}=\ket{\mu}\otimes\ket{m,n} \\[2pt] 
&T = T_{1}+T_{2}+(S_{1\alpha}+\gamma_{\alpha})\alpha^{\dagger}+(S_{1\beta}+\gamma_{\beta})\beta^{\dagger},\label{eq:Excited}
\end{split}
\end{equation}
where $\mu$ labels an electronic excitation while $n$ and $m$ are photonic excitations in $\alpha$ and $\beta$. The ground state energy is equal to
\begin{equation}
E = \bra{\textrm{HF},0,0}\textrm{exp}\left(-T\right)H\;\textrm{exp}\left(T\right)\ket{\textrm{HF},0,0}.
\label{eq:energy}
\end{equation}
The energy in Eq.~(\ref{eq:energy}) can be rewritten in terms of the cluster amplitudes and the two electron integrals $g_{pqrs}$ \cite{helgaker2014molecular} as
\begin{equation}
\begin{split}
E =&\; E_{\textrm{HF}} + \sum_{aibj}(t^{ab}_{ij}+t^{a}_{i}t^{b}_{j})(2g_{iajb}-g_{ibja})\\
+&\frac{\lambda}{\sqrt{2\tilde{\omega}}}\sum_{ai}\left[(\mathbf{p}\cdot\boldsymbol{\epsilon})e^{i\mathbf{k}\mathbf{r}}\right]_{ia}(s^{a}_{i\alpha} + \gamma_{\alpha} t^{a}_{i})\\
+&\frac{\lambda}{\sqrt{2\tilde{\omega}}}\sum_{ai}   \left[(\mathbf{p}\cdot\boldsymbol{\epsilon}^{*})e^{-i\mathbf{k}\mathbf{r}}\right]_{ia} (s^{a}_{i\beta} + \gamma_{\beta} t^{a}_{i}), 
\label{eq:Explicit_energy}
\end{split}
\end{equation}
where
\begin{equation}
E_{\textrm{HF}} = \bra{\textrm{HF}}H_{e}\ket{\textrm{HF}}.
\end{equation}
The electronic Hamiltonian, $H_{e}$, is given by the standard expression
\begin{equation}
 H_{e}= \sum_{pq}h_{pq}E_{pq}+\frac{1}{2}\sum_{pqrs}g_{pqrs}(E_{pq}E_{rs}-\delta_{qr}E_{ps}), 
\end{equation}
and $h_{pq}$ is the one electron integral \cite{helgaker2014molecular}.\\
Due to the non-hermiticity of the approach, the coupled cluster method can give complex energies when the Hamiltonian is complex, e.g. for molecules in external magnetic field \cite{stopkowicz2015coupled,hampe2017equation,piecuch1990coupled}. Since this case is implicitly contained in our framework, we must ensure that the energy in Eq.~(\ref{eq:Explicit_energy}) is real.
This condition is satisfied if $S_{1\alpha}$ and $S_{1\beta}$ (as well as $\gamma_{\alpha}$ and $\gamma_{\beta}$ ) fulfill the relations
\begin{equation}
s^{a}_{i\alpha} = -s^{a*}_{i\beta} \hspace{1.2cm} \gamma_{\alpha} = -\gamma^{*}_{\beta} 
\label{eq:Condition}
\end{equation}
for real $t^{a}_{i}$ and $t^{ab}_{ij}$.
These conditions are not accidental as the same relationship holds between the integrals $[(\mathbf{p}\cdot\boldsymbol{\epsilon}^{*})e^{-i\mathbf{k}\mathbf{r}}]_{pq}$ and $[(\mathbf{p}\cdot\boldsymbol{\epsilon})e^{i\mathbf{k}\mathbf{r}}]_{pq}$ when the orbitals $p$ and $q$ are real. Therefore, the conditions in Eq.(\ref{eq:Condition}) must arise from a symmetry in the Hamiltonian. \\
In the following, we demonstrate that Eq.~(\ref{eq:Condition}) holds in the exact limit  
\begin{align}
\ket{\psi} =&\;\textrm{exp}(\sum_{mn}(S_{m\alpha,n\beta}+\gamma_{m\alpha,n\beta})\alpha^{\dagger\;m}\beta^{\dagger\;n})\\
\times&\;\textrm{exp}(T)\ket{\textrm{HF},0,0},  \label{eq:QED-FCI}
\end{align}
with $T$ and $S_{m\alpha,n\beta}$ containing all the electronic excitations. The exact eigenfunction $\ket{\psi}$ satisfies
\begin{equation}
H\ket{\psi} = E\ket{\psi}    \label{eq:Schro}
\end{equation}
with a real $E$. Applying a unitary transformation
\begin{align}
V =& \textrm{exp}\left(i\pi (\alpha^{\dagger}\beta + \beta^{\dagger}\alpha)/2\right)\textrm{exp}\left(i\pi (\alpha^{\dagger}\alpha + \beta^{\dagger}\beta)/2\right)  \nonumber \\ 
&\hspace{2cm}V^{\dagger}\alpha V= -\beta \label{eq:Relation}\\
&\hspace{2cm}V^{\dagger}\beta V= -\alpha, \nonumber 
\end{align}
on both sides of Eq.~(\ref{eq:Schro}) we obtain
\begin{equation}
V^{\dagger}HV\;V^{\dagger}\ket{\psi} = EV^{\dagger}\ket{\psi}.    \label{eq:Schro_1}
\end{equation}
We can now use Eq.~(\ref{eq:Relation}) to exchange $\alpha$ with $-\beta$ and viceversa in both $H$ and the wave function.
Complex conjugation of the transformed Hamiltonian in Eq.(\ref{eq:Schro_1}) gives
\begin{equation}
\begin{split}
(V^{\dagger}HV)^{*} =& H_{e}+\tilde{\omega}\left(\alpha^{\dagger}\alpha+\beta^{\dagger}\beta+1\right)\\
+&\frac{\lambda}{\sqrt{2\tilde{\omega}}}\sum_{pq}\left[\left(\mathbf{p}_{i}\cdot\boldsymbol{\epsilon}\right)e^{i\mathbf{k}\mathbf{r}_{i}}\right]_{pq}E_{pq}(\alpha+\beta^{\dagger})\\
+&\frac{\lambda}{\sqrt{2\tilde{\omega}}}\sum_{pq}\left[(\mathbf{p}_{i}\cdot\boldsymbol{\epsilon}^{*})e^{-i\mathbf{k}\mathbf{r}_{i}}\right]_{pq}E_{pq}((\alpha^{\dagger}+\beta),    
\label{eq:Final_Squared}
\end{split}
\end{equation}
where we assume the orbitals are real. On the other hand, the transformed wave function is equals
\begin{equation}
\begin{split}
(V^{\dagger}\ket{\psi})^{*} =&\;\textrm{exp}(\sum_{mn}(S^{*}_{m\alpha,n\beta}+\gamma^{*}_{m\alpha,n\beta})\alpha^{\dagger\;n}\beta^{\dagger\;m}(-1)^{m+n})\\
\times&\;\textrm{exp}(T^{*})\ket{\textrm{HF},0,0}.  \label{eq:QED-FCI_1}
\end{split}
\end{equation}
Since the Hamiltonians in Eqs.~(\ref{eq:No_squared}) and (\ref{eq:Final_Squared})  are identical, the eigenfunctions in Eqs.~(\ref{eq:QED-FCI}) and (\ref{eq:QED-FCI_1}) must also be equal to each other. This implies that the amplitudes follow the relations
\begin{equation}
\begin{split}
 S_{m,n} =& (-1)^{m+n}S^{*}_{n,m} \\
 \gamma_{m,n} =& (-1)^{m+n}\gamma^{*}_{n,m} \\
 \textrm{T} =& \textrm{T}^{*},
\end{split}
\end{equation}
which proves the assertion.\\  
Considering the well know problems in using coupled cluster with complex Hamiltonians \cite{taube2006new,evangelista2019exact}, this represents a very interesting outcome. 
Our results show that when the imaginary part of the Hamiltonian is introduced through bosonic operators, such complications can be overcome by choosing an appropriate shape of the cluster operator. This is, to our knowledge, the first observation of these symmetries. In future work, we will explore the potential applications of these findings for molecules in static magnetic fields.
\section{Mechanism behind the cavity induced enantiomer discrimination}\label{sec:Mechanism}
In the section above we have shown that in a chiral cavity no symmetry of the Hamiltonian enforces two enantiomers to have the same energy. In this section we will focus more on the mechanism that allows circularly polarized fields to differentiate among optically active molecules. It is a well known fact that chiral molecules can be identified based on the direction they rotate the polarization plane of linearly polarized light \cite{barron2009molecular}. This property is in particular linked to the optical rotatory tensor $R^{0n}_{ij}$ \cite{craig1998molecular}
\begin{equation}
R^{0n}_{ij}=Re\bra{\psi_{0}}m_{i}\ket{\psi_{n}}\bra{\psi_{n}}p_{j}\ket{\psi_{0}},    \label{eq:optical_rotatory_tensor}
\end{equation}
where $m_{i}$ and $p_{j}$ denotes the $i$ and $j$ spacial components of the magnetic and electric dipole in velocity gauge, respectively. In Eq.~(\ref{eq:optical_rotatory_tensor}), $\psi_{0}$ and $\psi_{n}$ label electronic ground and excited state wave functions. For different enantiomers of the same molecule, the optical rotatory tensor changes its sign. 
The optical rotatory tensor is also linked to another critical property of chiral molecules, circular dichroism. This is the differential absorption of left- and right-handed light. Inside an optical cavity, the photonic and molecular degrees of freedom mix up to create a wave function that features both photons and electrons. The photonic contributions to the wave function are heavily influenced by the response properties of the molecular system in the cavity. This can be seen treating the electron-photon interaction perturbatively. For example, the excited state coefficient $C_{n}$ to first order equals
\begin{equation}
C^{(1)}_{n}= -\frac{\bra{\psi_{n}}V\ket{\psi_{0}}}{E_{n}+\tilde{\omega}-E_{\psi_{0}}},   \label{eq:perturbed}
\end{equation}
where we observe that the magnitude of the photonic contributions depends on the transition moment and $E_{n}-E_{o}$ is the transition energy between the ground and excited state.
Since enantiomers interact differently with circularly polarized light, they are dressed differently by RHCP and LHCP light. The energy differences between enantiomers can also be, in a first approximation, linked to the optical rotatory tensor. Indeed, the second order perturbation expression for the energy equals
\begin{widetext}
\begin{equation}
 E^{(2)}= -\frac{\lambda^{2}}{2\tilde{\omega}}\sum_{n}\left[\frac{\bra{\psi_{0}}\left(\mathbf{p}\cdot\boldsymbol{\epsilon}\right)e^{i\mathbf{k}\mathbf{r}}\ket{\psi_{n}}\bra{\psi_{n}}\left(\mathbf{p}\cdot\boldsymbol{\epsilon}^{*}\right)e^{-i\mathbf{k}\mathbf{r}}\ket{\psi_{0}}}{E_{n}+\tilde{\omega}-E_{0}}+\frac{\bra{\psi_{0}}\left(\mathbf{p}\cdot\boldsymbol{\epsilon}^{*}\right)e^{-i\mathbf{k}\mathbf{r}}\ket{\psi_{n}}\bra{\psi_{n}}\left(\mathbf{p}\cdot\boldsymbol{\epsilon}\right)e^{i\mathbf{k}\mathbf{r}}\ket{\psi_{0}}}{E_{n}+\tilde{\omega}-E_{0}}\right],
 \end{equation}
 which, after rotational averaging and expansion in $k$, becomes
 \begin{equation}
\langle E^{(2)} \rangle= -\frac{\lambda^{2}}{6\tilde{\omega}}\sum_{n}\sum^{3}_{i=1}\frac{\bra{\psi_{0}}p_{i}\ket{\psi_{n}}\bra{\psi_{n}}p_i\ket{\psi_{0}}}{E_{n}+\tilde{\omega}-E_{0}}-\frac{\lambda^{2}k}{6\tilde{\omega}}\sum_{n}\sum^{3}_{i=1}\frac{R^{0n}_{ii}}{E_{n}+\tilde{\omega}-E_{0}}+O(k^{2}).\label{eq:Second_order}
\end{equation}
\end{widetext}
Since the optical rotatory tensor contribution changes its sign depending on the chosen enantiomer, some chiral molecules will be stabilized by the circularly polarized field, while some others will be de-stabilized. We notice that the mixed electron-photon excited states provide the critical contribution for determining the field induced discrimination. The differential photon dressing is the cause of the field induced energy differences as the second order perturbation energy can be rewritten as
\begin{equation}
E^{(2)} = -\bra{\psi_{0}}V\sum_{n}C^{(1)}_{n}\ket{\psi_{n}}. \label{eq:dressing}
\end{equation}
In Eq.~(\ref{eq:dressing}), the terms on the left of the summation are identical for both enantiomers and it is therefore only the perturbed wave function, i.e. the photon dressing, that changes. The MC-QED-CCSD method includes the electron-photon excited states responsible for the effects discussed in Eq.~(\ref{eq:Second_order}) through the mixed electron-photon operators $S_{1\alpha}\alpha^{\dagger}$ and $S_{1\beta}\beta^{\dagger}$. Once applied on the reference state $\ket{HF,0,0}$, indeed, the cluster terms populate the full set of electon-photon excited states:
\begin{align}
\ket{\psi}& =\;\textrm{exp}\left((S_{1\alpha}+\gamma_{\alpha})\alpha^{\dagger}+(S_{1\beta}+\gamma_{\beta})\beta^{\dagger} \right)\nonumber\\
&\times\;\textrm{exp}\left(T_{1}+T_{2}\right)\ket{\textrm{HF}}\otimes\ket{0,0}\\
 & =\left(1+\left(S_{1\alpha}+\gamma_{\alpha}\right)\alpha^{\dagger}+\left(S_{1\beta}+\gamma_{\beta}\right)\beta^{\dagger}+...\right)\ket{\textrm{HF}}\nonumber \label{eq:QED-CCSD-S-MINIMAL}
\end{align}
with coefficients that depend on the amplitudes.
\section{Results}
\begin{figure}
    \centering
    \includegraphics[width=0.45\textwidth]{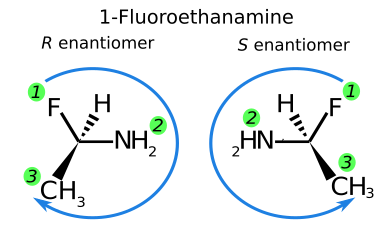}
    \caption{Identification of the absolute configuration of a chiral molecule. We first assign a priority number to each atom attached to the chiral center based on the atomic number (from largest to smallest). The structure is then rotated such that the lightest element (H in this case) is pointing backwards. Finally, an arrow is drawn from the highest to the lowest priority element. If the arrow rotates clockwise, we refer to the enantiomer as \textit{R}, otherwise we refer to the enantiomer as \textit{S}. This procedure provides a unique identification of the molecular structure \cite{clayden2012organic}}
    \label{fig:absolute}
\end{figure}
\begin{figure*}
    \centering
    \includegraphics[width=1.0\textwidth]{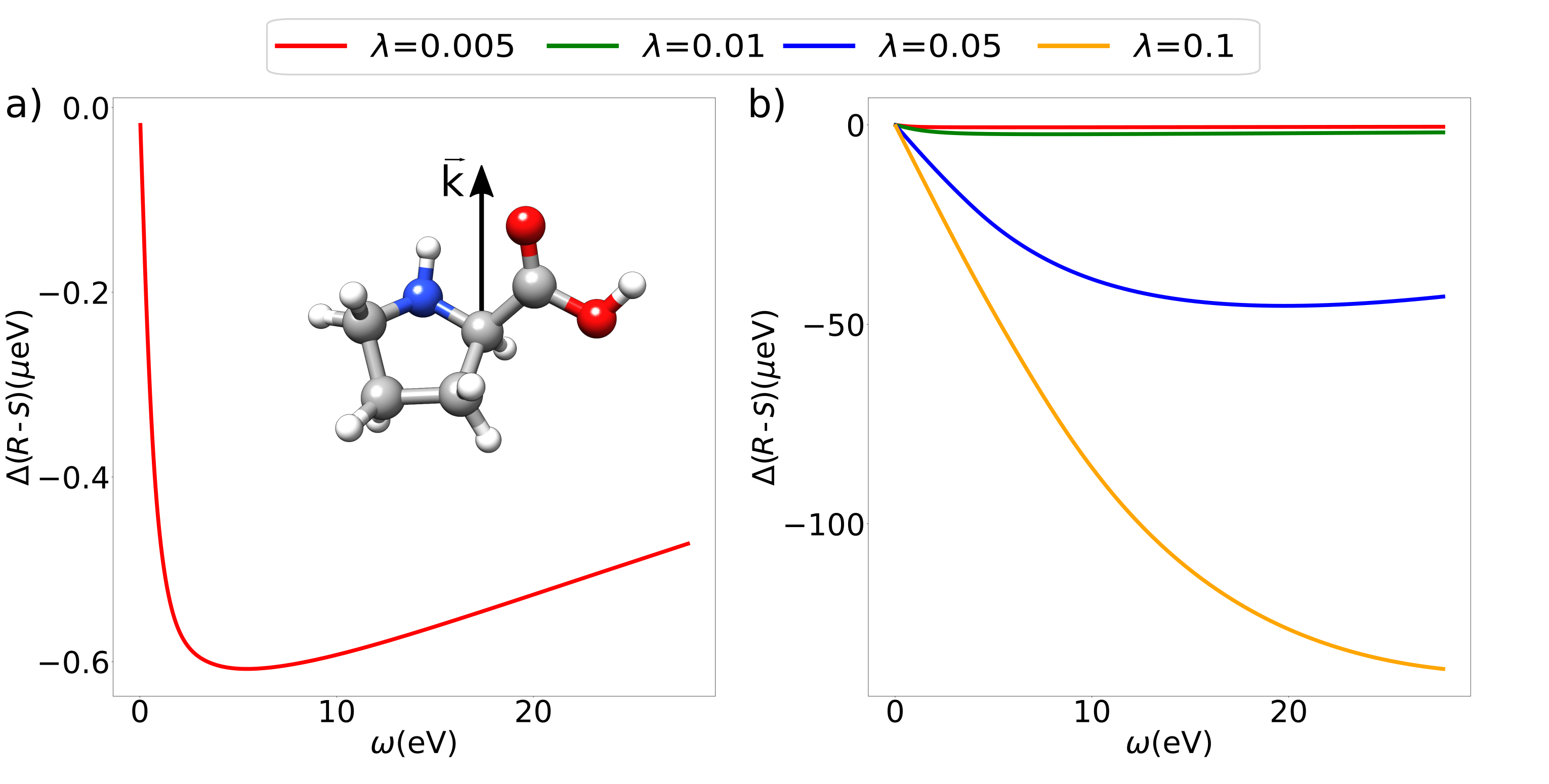}
    \caption{ a) Frequency dispersion of the discriminating power in a LHCP cavity at coupling $\lambda= 0.005$. The discriminating power reaches its minimum value around $\omega=4$ eV. b) Frequency dispersion of the discriminating power for different values of the coupling strength. The size of the discriminating power increases with the coupling but the qualitative shape of the dispersion is the same for all the analyzed cases. The minimum in $\Delta (\textit{R}-\textit{S})$ is shifted at higher frequencies as the coupling increases.}
    \label{fig:Dispersion_with_frequency}
\end{figure*}
\begin{figure*}
    \centering
    \includegraphics[width=1.0\textwidth]{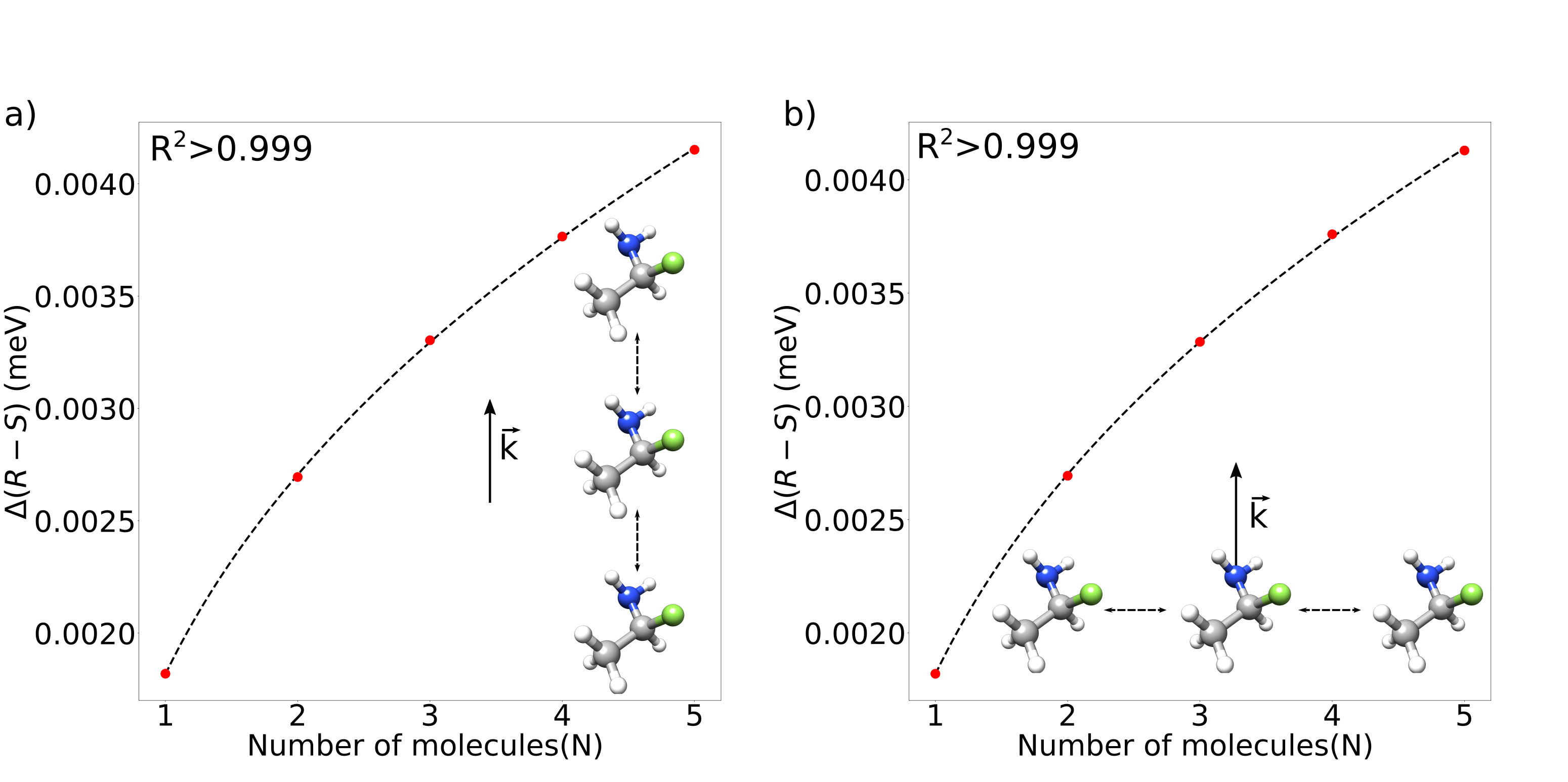}
    \caption{Effect of the number of molecules on the cavity discriminating power for \textit{R} and \textit{S} 1-fluoroethylamine displaced by 200 \AA\; along or perpendicularly to the wave vector. The fitting functions for the observed effects are: a) $\Delta\left(R-S\right)=0.0013+0.0015 \sqrt{N}-0.001/\sqrt{N}$; b) $\Delta\left(R-S\right)=0.0013+0.0014 \sqrt{N}-0.001/\sqrt{N}$.  }
    \label{fig:Collective}
\end{figure*}
\begin{figure}
    \centering
    \includegraphics[width=0.45\textwidth]{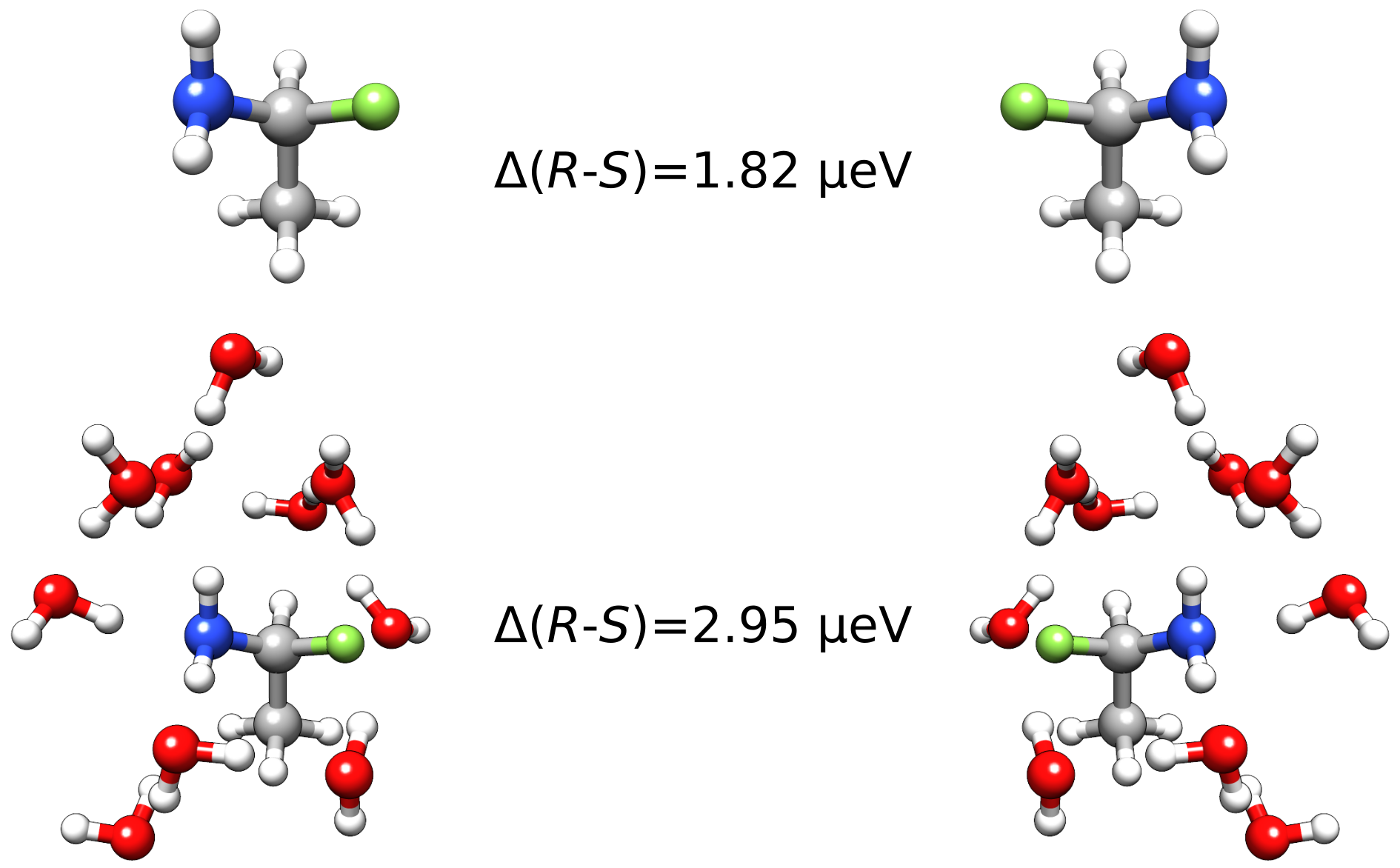}
    \caption{Solvent induced enhancement of the chiral discriminating effect. The solvent arranges in a chiral configuration around the chiral solute and the field interacts with a larger chiral system.}
    \label{fig:Solvent_role}
\end{figure}
In this section we apply the MC-QED-CCSD methodology on a set of chiral molecules interacting with a handedness-preserving cavity. Specifically, we perform a detailed quantitative analysis of the cavity-induced discrimination effects and we show that rotational spectra present enantioselective signatures in a chiral cavity.   
The calculations have been performed using a development version of the eT program \cite{folkestad2020t} using the cc-pVDZ basis set \cite{pritchard2019a,dunning1989a}. The molecular structures have been optimized with the ORCA software package \cite{neese2020orca} at the DFT-B3LYP/def2-SVP level\footnote{The geometries can be found in the repository https://doi.org/10.5281/zenodo.7035887
} \cite{weigend1998ri}  .
The basis set effects on the cavity calculations are discussed in Appendix C.
To uniquely define the enantiomers of a chiral molecule we use the absolute configuration notation \cite{clayden2012organic} explained in Fig.~\ref{fig:absolute}.

\subsection{Cavity induced discrimination power}
Strong coupling with circularly polarized electromagnetic fields creates energy differences between enantiomeric pairs. Since this is a field-induced effect, the cavity properties strongly influence the magnitude of the discriminating power.  
In Fig.~\ref{fig:Dispersion_with_frequency}, we show the frequency dispersion of the energy difference between \textit{R} and \textit{S} enantiomers of proline, $\Delta (\textit{R}-\textit{S})$, in a LHCP cavity. The sign of $\Delta (\textit{R}-\textit{S})$ remains the same for every frequency of the field and for every coupling value. The function changes sign when the circular polarization of the field is reversed. The magnitude of the discrimination is strongly affected by both $\omega$ and $\lambda$ and shows the same qualitative behaviour for all coupling values. Specifically, $\Delta (\textit{R}-\textit{S})$ is equal to zero for $\omega=0 \;\textrm{eV}$, then reaches a minimum (or a maximum) for an intermediate frequency and approaches zero again for $\omega\rightarrow\infty$. The curve shape can be explained using the previously developed theory. The discriminating power of the cavity approaches zero at small $\omega$ because the exponential $e^{ikr}\approx 1$ (see Section \ref{sec:Theoretical}). On the other hand, $\Delta (\textit{R}-\textit{S})$ also approaches zero for large cavity frequencies because the photonic component of the wave function becomes very small for very high energy fields (the coefficient of the first order perturbed wave function in Eq.~(\ref{eq:perturbed}) vanishes if $\omega$ approaches infinity). The position of the curve minimum is the most interesting feature of the dispersion curve. 
Photons with high frequency, i.e. large $\mathbf{k}$, have high discriminating power per photon because the second term in Eq.~(\ref{eq:Second_order}) becomes increasingly larger. However, since the population of the photonic states decreases when the field energy increases, an optimal point is found. This point depends on the coupling factor as shown in Fig.\ref{fig:Dispersion_with_frequency} b. In particular, we notice that the minimum location is shifted to higher frequencies when the coupling increases because $\lambda$ enhances the population of higher-energy and more discriminating photonic states. We point out that while the effect is still non-resonant, the optimal frequency value in Fig.\ref{fig:Dispersion_with_frequency} is system/cavity dependent.  \\
The chiral discriminating powers reported in Fig.\ref{fig:Dispersion_with_frequency} are very small for realistic values of the light matter coupling ($\lambda\leq0.05$ au$\;$). In particular, the field-induced energy differences are significantly smaller than the average thermal energy at room temperature. However, it is well-known that field-induced modifications can be amplified by many orders of magnitude through cooperation mechanisms between multiple molecules and field modes \cite{saez2018photon,ribeiro2021enhanced,li2021collective,mandal2022theory,sidler2020polaritonic,sidler2022perspective}. This class of phenomena is commonly called collective effects \cite{herrera2020molecular,mandal2022theory}. Even though the term collective effects is typically associated with excited states in the strong coupling community \cite{sidler2020polaritonic,cui2022collective}, optical cavities are known to induce long range correlation between molecular systems even in their ground state \cite{haugland2021intermolecular,riso2022characteristic,schafer2019modification,ruggenthaler2022understanding}. It is therefore interesting to investigate how the number of strongly coupled enantiomers influences the discrimination power of the cavity.
In Fig.\ref{fig:Collective} we plot the dependence of $\Delta\left(R-S\right)$ with respect to the number of chiral molecules in the cavity. Specifically, we perform calculations on a set of 1-fluoroethylamine (both \textit{R} and \textit{S}) separated by 200 \AA\hspace{0.5mm} either along the wave vector direction, Fig.\ref{fig:Collective} a, or perpendicularly to $\mathbf{k}$, Fig.\ref{fig:Collective} b. The coupling is fixed to $\lambda=0.05\;au$ while the frequency of the cavity is equal to $1.36$ eV. In both directions the discriminating power of the cavity is enhanced as the number of chiral centers increases. The dispersions in Fig.\ref{fig:Collective} are slower than linear and the effect is dominated by the square root of the number of enantiomers in the cavity. This trend is in line with the behaviour shown in Ref.~\cite{schafer2022chiral}, see Appendix F for additional details. We note, however, that the enantiomeric discrimination discussed here is a non-resonant effect, showing the typical ground state  dependence from the cavity parameters, i.e. $\Delta\left(R-S\right)\propto O(\lambda^{2})$. The fitting functions in Fig.\ref{fig:Collective} have been chosen considering that the main energy contribution from the light-matter interaction term scales as $\sqrt{N}$.
Moreover, for $N_{e}\lambda^{2}\gg\omega^{2}$ we have:
\begin{equation}
\frac{\lambda}{\sqrt{2\sqrt{\omega^{2}+N_{e}\lambda^{2}}}} \approx \frac{1}{\sqrt{2\lambda\sqrt{N_{e}}}}\left(\lambda-\frac{\omega^{2}}{2N_{e}}\right), \label{eq:expansion}
\end{equation}
which justifies the second contribution in the fit.
The field induced energy differences are slightly stronger when molecules are displaced along $\mathbf{k}$.
Although the effect becomes infinitely large as the number of molecules increases, the cavity-induced differences are not extensive. Our results are in disagreement with those reported in Ref.~\onlinecite{galego2015cavity}, where the authors show that ground state properties do not depend on the number of molecules in the cavity. This disagreement, in our opinion, is due to the inclusion of the squared field term, $A^{2}$, in the Hamiltonian, see Eq.(6). This contribution, which is on par with the dipole self-energy in the formalism of Ref.~\onlinecite{galego2015cavity}, changes the effective frequency of the cavity from $\omega$ to $\tilde{\omega}$ introducing a dependence on the number of molecules also for ground state properties. Differently from resonant effects, therefore, it is not the effective coupling with the field that changes, but the frequency. Since the frequency variations are proportional to $\lambda^{2}$, neglecting such terms becomes more and more exact as the coupling decreases and the approach in Ref.~\onlinecite{galego2015cavity} is obtained.
If the chiral molecule is dissolved in a liquid, the solvent too plays a significant role in enhancing the field-discriminating power. Indeed, when the solute is chiral the solvent arranges in a chiral structure itself at least in the first few solvation shells. Under these conditions the field interacts with a significantly larger chiral system increasing the discriminating effect. For example, in Fig.~\ref{fig:Solvent_role} we show that when 10 water molecules solvate 1-fluoroethylamine, the cavity effect almost doubles. We envision that the solvent enhancement should become even more intense if additional chiral molecules are added in the solution or if a chiral solvent is used in the first place. This topic will be the subject of a future publication. Additional anisotropy factors, like magnetic fields or pulses of external circularly polarized light, could also enhance the cavity discrimination similarly to what happens for MCD spectra \cite{sun2019ab,stephens1970theory,stephens1974magnetic}.
All the factors discussed above should increase the field-induced energy differences in chiral cavities, potentially to the kJ/mol range.

\subsection{Rotational spectra in chiral cavities}
\begin{figure}
    \centering
    \includegraphics[width=0.45\textwidth]{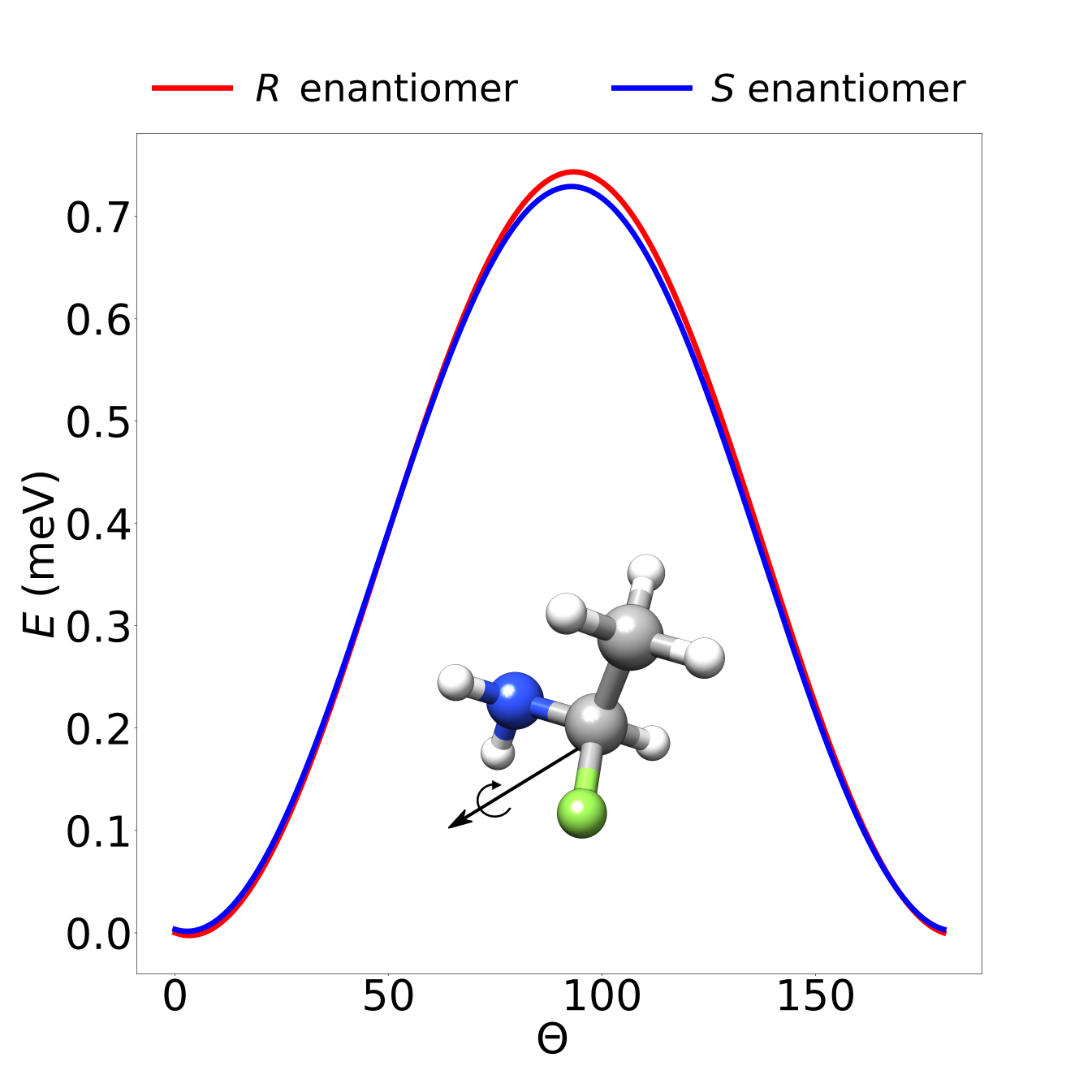}
    \caption{Orientational effects for \textit{R} and \textit{S} enantiomers of 1-fluoroethylamine upon rotation around an arbitrary axis shown above of an angle $\theta$.}
    \label{fig:Orientational}
\end{figure}
\begin{figure}
    \centering
    \includegraphics[width=0.4\textwidth]{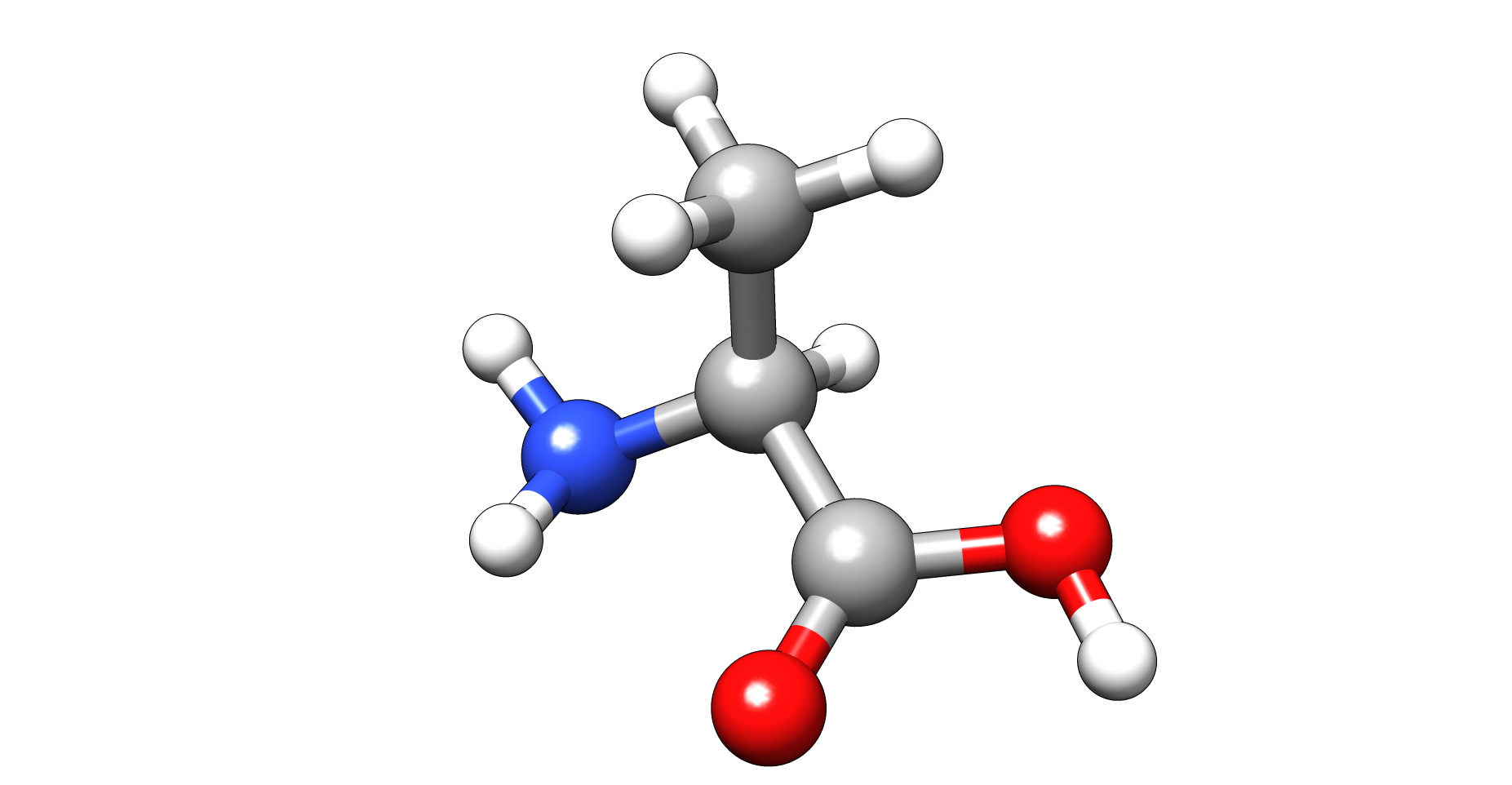}
    \caption{\textit{S} enantiomer of the alanine molecule.}
    \label{fig:alanine}
\end{figure}
\begin{figure}
    \centering
    \includegraphics[width=0.4\textwidth]{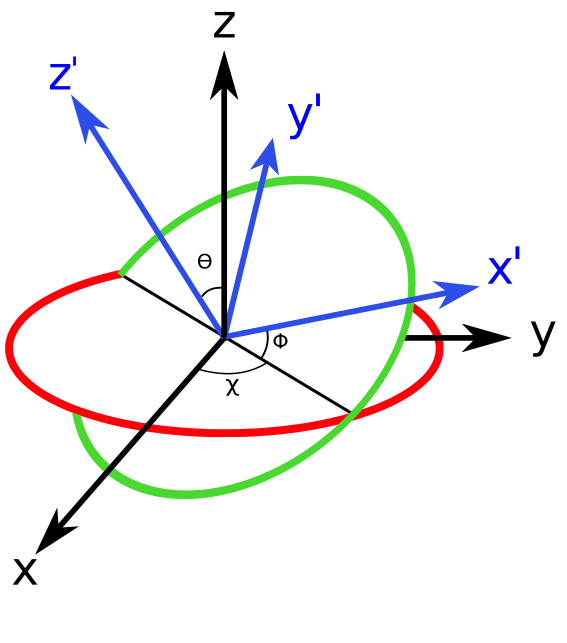}
    \caption{Pictorial representation of the Euler angles}
    \label{fig:Euler}
\end{figure}

\begin{figure*}
    \centering
    \includegraphics[width=1.0\textwidth]{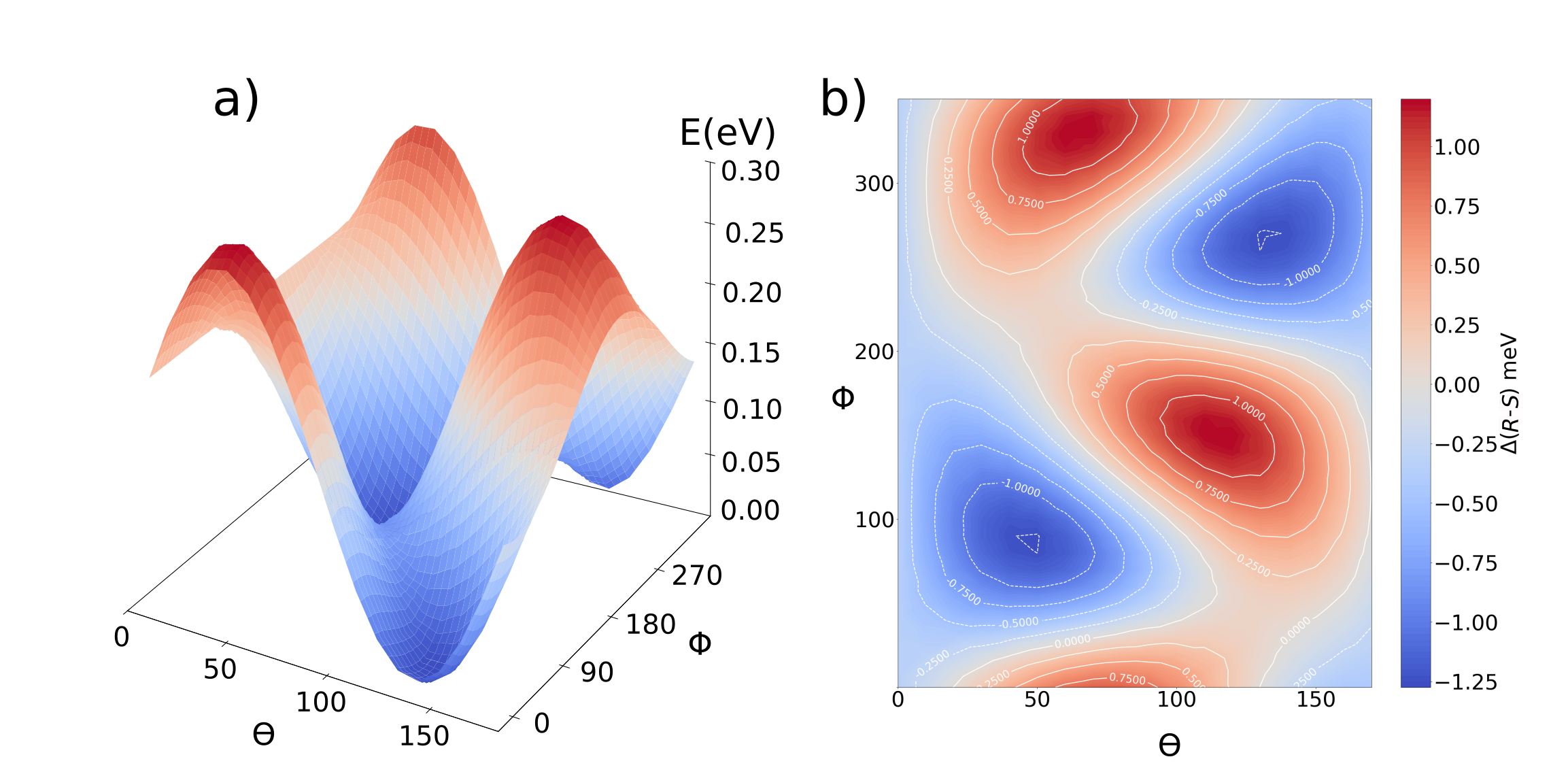}
    \caption{a) Rotational surface for \textit{S} alanine in a LCHP cavity. The surface is not flat, instead some configurations are stabilized by the field. b) Difference in the rotational surfaces of \textit{R} and \textit{S} alanine. We observe that the energy difference does not have a constant sign.}
    \label{fig:Rotational}
\end{figure*}
\begin{figure*}
    \centering
    \includegraphics[width=1.0\textwidth]{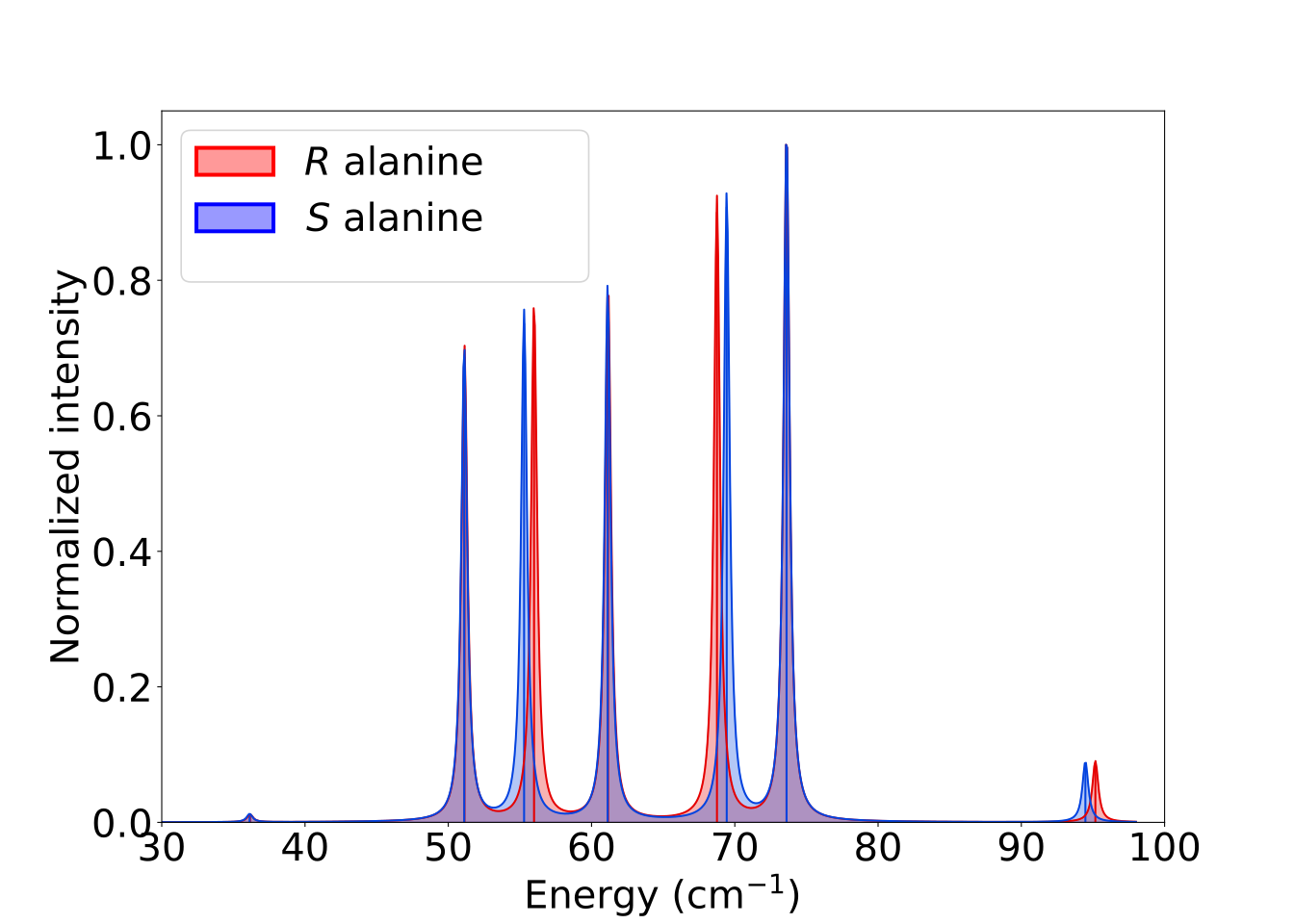}
    \caption{Rotational spectra of \textit{R} and \textit{S} alanine inside an LCHP cavity.}
    \label{fig:Peaks}
\end{figure*}

Inside an optical cavity, the molecular energy is highly dependent on the molecular orientation because the quantization direction for the cavity field, $\mathbf{k}$, naturally introduces a spacial anisotropy \cite{haugland2021intermolecular}. For example, in Fig.~\ref{fig:Orientational} we show how the energy of 1-fluoroethylamine in a LHCP field changes upon rotation of the molecule around an arbitrary axis. The photons, indeed, dress molecules differently depending on their orientation. Using the concepts discussed in Sec.~\ref{sec:Mechanism}, this follows from the idea that a molecule does not interact with light in the same way in all orientations. Some configurations are therefore stabilized more than others by the cavity resulting in preferential orientations of the system. This should have a significant effect on the rotational spectra of the system in the cavity.
We point out that we expect modifications in the rotational spectrum even for strong coupling to linearly polarized fields. However, different enantiomers would surely exhibit the same rotational levels. Inside, a chiral cavity, instead, due to the differential photon dressing of the enantiomers, the field induced orientational stabilization should differ for the two enantiomers. In particular, the orientational effects reported in Fig.~\ref{fig:Orientational} for the \textit{R} and \textit{S} enantiomers are not exactly the same, confirming that the rotational surfaces differ for the two mirror images. These variations should induce enantioselective signatures in the rotational spectra. In this section we compute the cavity-induced orientational effects on the two enantiomers of alanine (see Fig.\ref{fig:alanine}) and the relative rotational spectra. All the possible molecular orientations are obtained varying the three Euler angles $\phi$, $\theta$ and $\chi$ describing a rotation around $z$, a rotation around $x$ and a rotation on around $z$ again, respectively (see Fig.\ref{fig:Euler}). The energy is invariant with respect to $\chi$ for symmetry reasons. The Hamiltonian used to describe the nuclear motion is
\begin{equation}
\begin{split}
H =& \sum_{I}\frac{p^{2}_{I}}{2M_{I}}+V(R),
\label{eq:Nuclear_motion}
\end{split}
\end{equation}
where $V(R)$ is the potential energy surface obtained perfoming a MC-QED-CCSD calculation for every orientation of the molecule. The interactions between the nuclei and the field in Eq.~(\ref{eq:Nuclear_motion})
are neglected because the nuclear motion is much slower than the electronic one. Once the motion of the center of mass has been removed, the kinetic contribution in Eq.~(\ref{eq:Nuclear_motion}) can be split into a vibrational $H_{vib}$ and a rotational part $H_{rot}$ (the mixed rovibrational contribution will be neglected in this analysis). The Hamiltonian in Eq.~(\ref{eq:Nuclear_motion}) is therefore rewritten as:
\begin{equation}
\begin{split}
&H =H_{rot}+H_{vib}\\
H_{rot}=& \frac{J^{2}_{\xi}}{2I_{\xi}}+\frac{J^{2}_{\eta}}{2I_{\eta}}+\frac{J^{2}_{\zeta}}{2I_{\zeta}}+V_{rot}(\theta,\phi)
\label{eq:Rigid_Rotator}
\end{split}
\end{equation}
where $\xi$, $\eta$ and $\zeta$ are the principal axes of inertia of the molecule, treated as a rigid body \cite{landau2013quantum}. A detailed discussion of how to compute the rotational energy levels from Eq.~(\ref{eq:Rigid_Rotator}) is reported in Appendix D. The rotational potential energy surface, $V_{rot}(\theta,\phi)$, for the S enantiomer of alanine in a LHCP cavity is shown in Fig.~\ref{fig:Rotational} a. Differently from the no-cavity case, the surface is not flat and presents two maxima and two minima: the most destabilized and stabilized configurations, respectively. The difference in the orientational effects for the two enantiomers ($\Delta(\textit{R}-\textit{S})$) is plotted in Fig.~\ref{fig:Rotational} b. In particular, the results shown in Fig.~\ref{fig:Rotational} are obtained summing up the cavity orientational effects obtained from the first 20 modes of a chiral cavity with fundamental excitation at $2.7$ eV. The coupling has been set to $\lambda = 0.05$ au. We note that in Fig.~\ref{fig:Rotational} b, $\Delta(\textit{R}-\textit{S})$ does not have a constant sign implying that the surfaces are not just shifted by a constant. Specifically, the potential of the \textit{S} enantiomer has higher maxima and lower minima when compared to the surface obtained for the \textit{R} enantiomer. The rotational spectra of the two mirror images are shown in Fig.~\ref{fig:Peaks}. Due to the orientational effects in the cavity, the peaks are shifted to higher energies than standard rotational spectroscopy. We highlight that \textit{R} and \textit{S} spectra are not identical. While the intensities are mostly unchanged, the peak positions are slightly different with the largest modifications observed on the signals around 55, 70 and 100 cm$^{-1}$. The shifts are not always in the same direction and we observe that the \textit{R} enantiomer spectrum has a lower excitation around 70 cm$^{-1}$ while the \textit{S} enantiomer spectrum shows lower transitions at 55 and 100 cm$^{-1}$. Differences in the peak positions are on the order of 0.5 cm$^{-1}$, which is still large enough to be detected experimentally. However, the shifts can be enhanced by solvent and collective effects as discussed in the previous section.
Together with the data shown before, these results confirm that chiral cavities can be used to create energy differences between enantiomers and that they induce enantiospecific shifts in their rotational spectra. Observation of the enantiospecific signatures in rotational spectroscopy would be a clear experimental proof of the effects described in this paper. We point out, however, that dissipative channels might be open for polaritonic states and that they might affect the spectral resolution achievable experimentally. The eventual broadening of the peaks, indeed, has not been included in Fig.\ref{fig:Peaks}. 
We envision that the cavity induced discrimination effects described in this section might be used to improve processes where enantiomers need to be separated \cite{vu2022enhanced}. As discussed in the introduction, this is a critical process for non enantioselective synthesis methods where the differentiation among enantiomers has to be performed after a 50/50 \% mixture of the two mirror images has been formed. Moreover, in future works we will test weather the use of chiral cavities can be used to induce enantioselectivity in chemical reactions.    
\section{Conclusion}
In this work, we present the first \textit{ab initio} framework to model strong coupling between molecules and circularly polarized light. We show that the simplest theoretical approach to properly describe these systems requires the inclusion of two cavity modes to ensure the correct field symmetry. We also discuss how inclusion of the beyond dipole contributions is critical to capture the chiral nature of the field. Our implementation is, to our knowledge, the first report of an \textit{ab initio} QED approach where the full field shape has been used. This is a significant improvement over previous methodologies since the inclusion of the full field shape allows for the treatment of any kind of cavity dimension and field frequency. This choice also solves serious issues with beyond dipole QED approaches, e.g. multipolar expansion, that introduces an expansion point dependence in the results. We investigate the formal properties of the chiral cavity Hamiltonian and argue that, using circularly polarized electromagnetic fields, it is indeed possible to discriminate between the two mirror images of a chiral molecules. To perform numerical simulations on realistic systems we develop a complex QED coupled cluster approach. This is a new critical development as the non-hermiticity of the approach makes it challenging to deal with complex Hamiltonians without unphysical complex energies. We show that for QED methods the energy remains real if an appropriate form of the cluster operator is employed.
Our results demonstrate that chiral cavities create energy differences between enantiomers. The sign of the chiral discrimination does not depend on the frequency of the field or the coupling strength. Instead, it is only affected by the circular polarization of the cavity. This is an essential observation for future experimental applications as the effects are stable for very large variations of the cavity parameters.

The dependence of the discrimination on the number of strongly coupled chiral molecules has also been investigated. In particular, we observe an increase in the discriminating power with the number of chiral systems in the cavity. The solvent also enhances the enantioselective effect by creating a chiral solvation shell around the chiral solute. Finally, we demonstrate that enantiomers do not have the same rotational spectra in chiral cavities. Specifically, the circularly polarized field induces enantiospecific shifts in the peak positions. Our results suggest that interesting phenomena should be observable when molecules are placed in chiral cavities. Future developments will tackle the calculation of excited states for chiral molecules in chiral cavities \cite{schafer2022chiral,sun2022polariton}. Moreover, future investigations will deal with the possibility to use circularly polarized fields to induce enantioselectivity in chemical reactions\cite{ke2022can,baranov2022towards,schafer2022chiral}.
We believe that, together with other reported findings \cite{vu2022enhanced}, our results provide the necessary motivation for further investigations of strong coupling in enantiomeric separation. 
\subsection*{Acknowledgements}
We acknowledge Tor S. Haugland for insightful discussions. R.R.R and H.K. acknowledge funding from the Research Council of Norway through FRINATEK Project No. 275506. We acknowledge computing resources through UNINETT Sigma2—the National Infrastructure for High Performance Computing and Data Storage in Norway, through Project No. NN2962k. This work has received funding from the European Research Council (ERC) under the European Union’s Horizon 2020 Research and Innovation Programme (grant agreement No. 101020016).
\subsection*{Appendix A: Zero-point energy and non size-extensive effects}
The $\theta$ value in the Bogoljubov transformation, Eq.~(\ref{eq:Rotation}), has been chosen to remove the quadratic contribution in Eq.(\ref{eq:Twomodes}). The same $\theta$ value is obtained when minimizing the zero-point energy. Indeed, after the Bogoljubov transformation, the zero-point energy becomes equal to
\begin{equation}
\mathscr{E}(\theta,N_{e}) = \frac{N_{e}\lambda^{2}}{2\omega}\left(\cosh\theta-\sinh\theta\right)^{2}+2\omega\sinh^{2}\theta+\frac{\omega}{2},    
\end{equation}
which has a minimum for 
\begin{equation}
\tanh2\theta = \frac{\lambda^{2} N_{e}}{\lambda^{2}N_{e}+2\omega^{2}}.
\label{eq:theta}
\end{equation}
Using Eq.(\ref{eq:theta}), the zero-point energy becomes equal to:
\begin{equation}
\mathscr{E}_{0}(N_{e}) = \frac{1}{2}\sqrt{\omega^{2}+N_{e}\lambda^{2}}=\frac{\omega}{2}\sqrt{1+\frac{N_{e}\lambda^{2}}{\omega^{2}}},  
\label{eq:Energy}
\end{equation}
which is non size-extensive due to the square root dependence on the number of molecules. The non size-extensize contribution to the zero-point energy equals
\begin{equation}
E_{nse}=\mathscr{E}_{0}(N_{e})-N_{e}\mathscr{E}_{0}(1)+\frac{(N_{e}-1)\omega}{2}, \label{eq:Non_size}
\end{equation}
where the last term has been added to avoid overcounting of the cavity zero-point energy.
When $\frac{N_{e}\lambda^{2}}{\omega^{2}}<<1$, Eq.~(\ref{eq:Non_size}) can be expanded in a Taylor series leading to
\begin{equation}
E_{nse}\approx-\frac{N_{e}(N_{e}-1)\lambda^{4}}{8\omega^{3}},
\end{equation}
which shows the same $N_{e}(N_{e}-1)$ behaviour already reported in Ref.\cite{haugland2021intermolecular}.
\subsection*{Appendix B: Bogoljubov transformation for the multimode case}
The transformation in Eq.(\ref{eq:Rotation}) is also useful when more modes are included in the minimal coupling Hamiltonian:
\begingroup
\allowdisplaybreaks
\begin{align}
H =& \sum_{i}\frac{\mathbf{p}^{2}_{i}}{2}+\sum_{k>0}\omega_{k}\left(b^{\dagger}_{k}b_{k}+b^{\dagger}_{-k}b_{-k}+1\right)\nonumber\\
+&\sum_{i>j}\frac{1}{\left|r_{i}-r_{j}\right|}+\sum_{I>J}\frac{Z_{I}Z_{J}}{\left|R_{I}-R_{J}\right|}-\sum_{i, I}\frac{Z_{I}}{\left|R_{I}-r_{i}\right|}\nonumber\\
+&\lambda\sum_{i}\sum_{k>0}\left(\mathbf{p}_{i}\cdot\boldsymbol{\epsilon}_{\pm}\right)e^{i\mathbf{k}\mathbf{r}_{i}}\frac{(b_{k}+b^{\dagger}_{-k})}{\sqrt{2\omega_{k}}}\nonumber\\
+&\lambda\sum_{i}\sum_{k>0}(\mathbf{p}_{i}\cdot\boldsymbol{\epsilon^{*}}_{\pm})e^{-i\mathbf{k}\mathbf{r}_{i}}\frac{(b^{\dagger}_{k}+b_{-k})}{\sqrt{2\omega_{k}}}\label{eq:Multiple}\\
+&\sum_{k>0}\frac{N_{e}\lambda^{2}}{2\omega_{k}}\left(b_{k}+b^{\dagger}_{-k}\right)\left(b_{-k}+b^{\dagger}_{k}\right)\nonumber\\
+&\sum_{ik\neq k^{\prime}>0}\frac{N_{e}\lambda^{2}e^{i(\mathbf{k}-\mathbf{k}^{\prime})\mathbf{r}_{i}}}{2\sqrt{\omega_{k}\omega_{k^{\prime}}}}\left(b_{k}+b^{\dagger}_{-k}\right)\left(b_{-k^{\prime}}+b^{\dagger}_{k^{\prime}}\right).\nonumber
\end{align}
\endgroup
In this case the squared term couples different field modes. Due to the presence of electronic operators, the quadratic contributions in the field cannot be fully reabsorbed using a unitary bosonic transformation similar to Eq.~(\ref{eq:Rotation}). However, a product of those rotations leads to
\begin{equation}
U = \prod_{k>0}\textrm{exp}\left[\theta_{k}\left(b^{\dagger}_{k}b^{\dagger}_{-k}-b_{k}b_{-k}\right)\right],
\label{eq:Prod_Rotation}
\end{equation}
which can be used to reabsorb the purely photonic terms of Eq.~(\ref{eq:Multiple}) leading to
\begin{equation}
\begin{split}
H =& \sum_{i}\frac{\mathbf{p}^{2}_{i}}{2}+\sum_{k}\tilde{\omega}_{k}\left(\alpha^{\dagger}_{k}\alpha_{k}+\beta^{\dagger}_{k}\beta_{k}+1\right)\\
+&\sum_{i>j}\frac{1}{\left|r_{i}-r_{j}\right|}+\sum_{I>J}\frac{Z_{I}Z_{J}}{\left|R_{I}-R_{J}\right|}-\sum_{i, I}\frac{Z_{I}}{\left|R_{I}-r_{i}\right|}\\
+&\lambda\sum_{i}\sum_{k}\left(\mathbf{p}_{i}\cdot\boldsymbol{\epsilon}_{\pm}\right)e^{i\mathbf{k}\mathbf{r}_{i}}\frac{(\alpha_{k}+\beta^{\dagger}_{k})}{\sqrt{2\tilde{\omega}_{k}}}\\
+&\lambda\sum_{i}\sum_{k}(\mathbf{p}_{i}\cdot\boldsymbol{\epsilon^{*}}_{\pm})e^{-i\mathbf{k}\mathbf{r}_{i}}\frac{(\alpha^{\dagger}_{k}+\beta_{k})}{\sqrt{2\tilde{\omega}_{k}}}\\
+&\sum_{i\mathbf{k}\neq\mathbf{k}^{\prime}}\frac{N_{e}\lambda^{2}e^{i(\mathbf{k}-\mathbf{k}^{\prime})\mathbf{r}_{i}}}{2\sqrt{\tilde{\omega}_{k}\tilde{\omega}_{k^{\prime}}}}\left(\alpha_{k}+\beta^{\dagger}_{k}\right)\left(\beta_{k^{\prime}}+\alpha^{\dagger}_{k^{\prime}}\right).
\label{eq:Multiple_v2}
\end{split}
\end{equation}
Once more, the frequencies have been redefined as $\tilde{\omega}_{k}=\sqrt{\omega^{2}_{k}+N_{e}\lambda^{2}}$.
\subsection*{Appendix C: Basis set effects in QED calculations}
\begin{figure}
    \centering
    \includegraphics[width=0.5\textwidth]{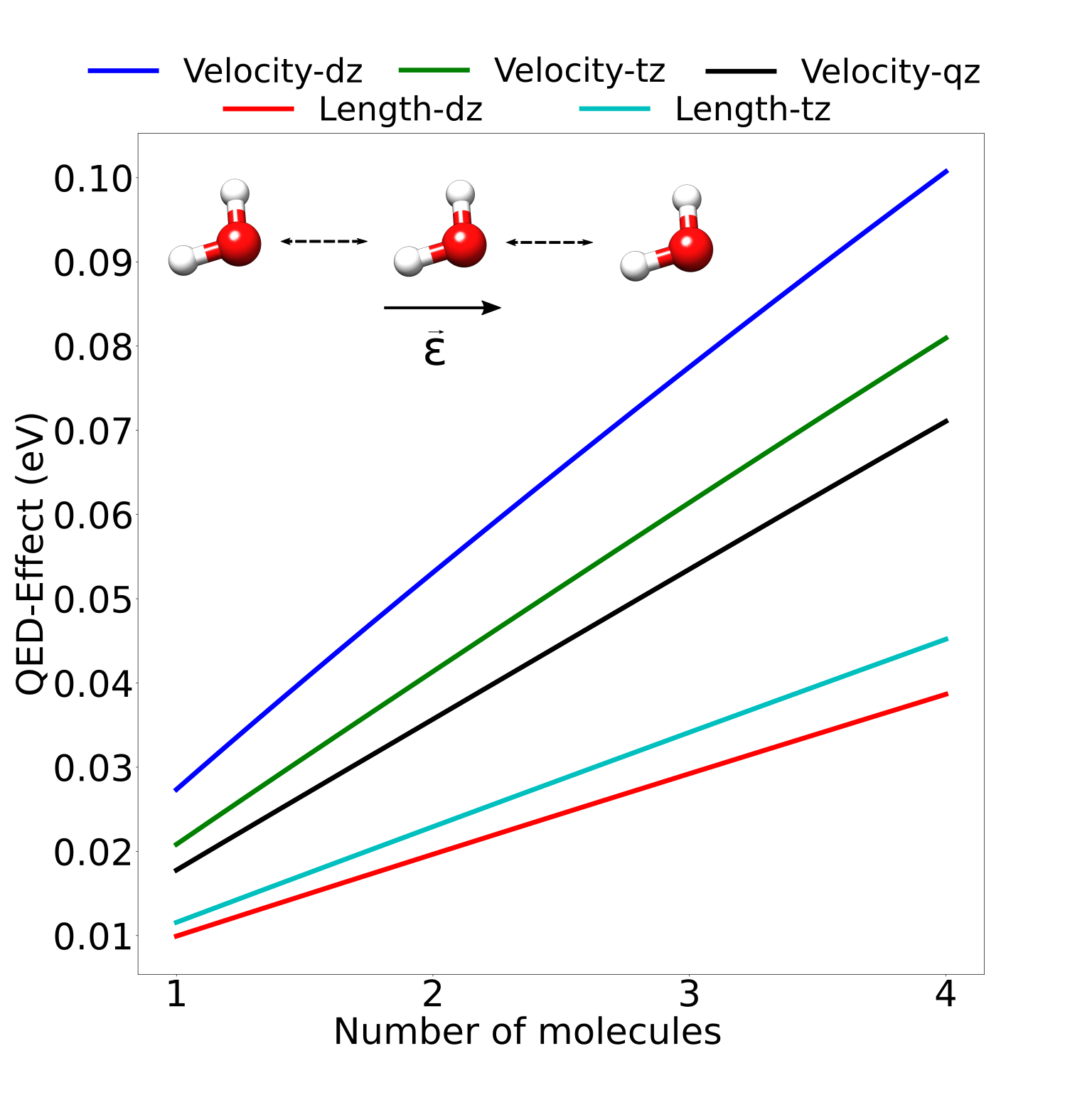}
    \caption{Dispersion of the QED-Effects for a system of water molecules displaced of 10 \AA\hspace{0.5mm} from each other. The coupling is equal to 0.1 au while the frequency is equal to $13.6$ eV.}
    \label{fig:QED_effects}
\end{figure}
Usually, the minimal coupling Hamiltonian is not employed to describe the interaction between light and matter, as it involves the complex momentum operator. Instead, it is customary to perform calculations using the unitarily transformed version of the minimal coupling Hamiltonian, called the PZW Hamiltonian. When the dipole approximation is adopted, the PZW Hamiltonian is equal to
\begin{equation}
\begin{split}
H_{\textrm{LG}}=& H_{e}+\lambda\sqrt{\frac{\omega}{2}}\sum_{i}(\mathbf{d}_{i}\cdot\boldsymbol{\epsilon})(b+b^{\dagger}) \\
+& \frac{\lambda^{2}}{2}\sum_{ij}(\mathbf{d}_{i}\cdot\boldsymbol{\epsilon})(\mathbf{d}_{j}\cdot\boldsymbol{\epsilon})+\omega b^{\dagger}b,  \label{eq:Length}  
\end{split}
\end{equation}
where $H_{e}$ is the standard electronic Hamiltonian and $\mathbf{d}_{i}$ is the molecular dipole operator. 
This Hamiltonian is referred to as length gauge form (LG). Extension of Eq.~(\ref{eq:Length}) beyond the dipole approximation is a relatively easy task if a multipole expansion is applied. However, this approach breaks the origin invariance of the problem. This is the reason why we have chosen a modified minimal coupling Hamiltonian to describe strongly coupled systems in chiral cavities. The two approaches (multipole expansion and minimal coupling) are completely equivalent in theory, but the basis truncation in real calculations makes them differ. 

In this section we therefore compare the QED-effects obtained using the minimal coupling Hamiltonian in the dipole approximation, also called the velocity-gauge Hamiltonian (VG), and the length-gauge Hamiltonian in Eq.~(\ref{eq:Length}).
In Fig.\ref{fig:QED_effects} we show the dispersion of the QED effects for an increasing number of water molecules spaced by 10 \AA\hspace{0.5mm} along the $\boldsymbol{\epsilon}$ direction. As expected, the results obtained using the different gauges are not the same. In particular, the velocity gauge consistently predicts larger field contributions than length gauge. However, the agreement improves as the basis set is enlarged. While the QED-effects computed using velocity gauge decrease when the basis is enlarged, the effects computed using length gauge increase. From Fig.\ref{fig:QED_effects} we infer that both gauges converge in the complete basis limit, one from above, and the other from below. Because of the non-variational character of the coupled cluster approach,
gauge invariance is not necessarily reached in the complete basis unless the full set of excitations are included \cite{pedersen1997coupled}. However, as shown in previous works \cite{pedersen1998gauge}, reasonable agreement between the two gauges can be reached with large basis sets and the two frameworks describe the same qualitative effects regardless of the basis size.
\subsection*{Appendix D: Calculation of the rotational spectrum}
The molecular rotational spectra shown in this paper are obtained treating the molecule as an asymmetrical top, following the theory in Ref. \cite{landau2013quantum}. In particular, the rotational Hamiltonian, $H_{\textrm{rot}}$ is equal to
\begin{equation}
\begin{split}
H_{rot} =& \frac{1}{4}\left(\mathbf{J}^{2}-J^{2}_{\zeta}\right)\left(\frac{1}{I_{\xi}}+\frac{1}{I_{\eta}}\right)+\frac{J^{2}_{\zeta}}{2I_{\zeta}}\\
 +&\frac{1}{8}\left(J^{2}_{+}+J^{2}_{-}\right)\left(\frac{1}{I_{\xi}}-\frac{1}{I_{\eta}}\right)+V(\theta,\phi), 
 \label{eq:New_Rotational}
\end{split}
\end{equation}
expressed in terms of angular momentum operators $J_{+}$ and $J_{-}$.    
The angular momentum component $\textrm{M}$ along the fixed $z$ axis is a good quantum number for the stationary states of Eq.~(\ref{eq:New_Rotational}).
The eigenvalues and eigenvectors of Eq.(\ref{eq:New_Rotational}) are obtained solving the eigenvalue problem 
\begin{equation}
\sum_{Jk^{\prime}}(\bra{JMk}H_{rot}\ket{J^{\prime}Mk^{\prime}}-E\delta_{kk^{\prime}})c_{J^{\prime}k^{\prime}}=0,    
\end{equation}
where $\{\ket{JMk}\}$ are the eigenfunctions of the symmetric top problem\cite{landau2013quantum}
\begin{equation}
D^{j}_{Mk}(\chi, \theta, \phi) = \textrm{exp}(iM\chi)d^{j}_{Mn}\textrm{exp}(ik\phi).    
\end{equation}
While the matrix elements for the $J$ operators are obtained using the angular momentum properties, the contribution arising from the cavity-induced potential is more difficult to account for. To include the contribution from the cavity potential we perform a discrete Fourier transform in $\theta$ and $\phi$
\begin{equation}
V(\theta,\phi) \approx \sum_{ln}V_{ln}e^{il\theta}e^{in\phi},
\end{equation}
where we have only retained the $V_{ln}$ such that $\abs{V_{ln}}>0.001$. The diagonalization is performed in \textit{Mathematica} for quantum number M = 0, 1, 2 and states with $J\leq 4$. The intensities in the spectrum have been computed using the oscillator strength
\begin{equation}
f_{eg} = \frac{2}{3}(E_{e}-E_{g})\sum_{n=x,y,z}\abs{\bra{\psi_{e}}d_{n}\ket{\psi_{g}}}^{2},    
\end{equation}
where $\psi_{g}$ and $\psi_{e}$ are the ground and excited state wave functions, respectively, with energies $E_{g}$ and $E_{e}$. 
\begin{figure*}
    \centering
    \includegraphics[width=0.78\textwidth]{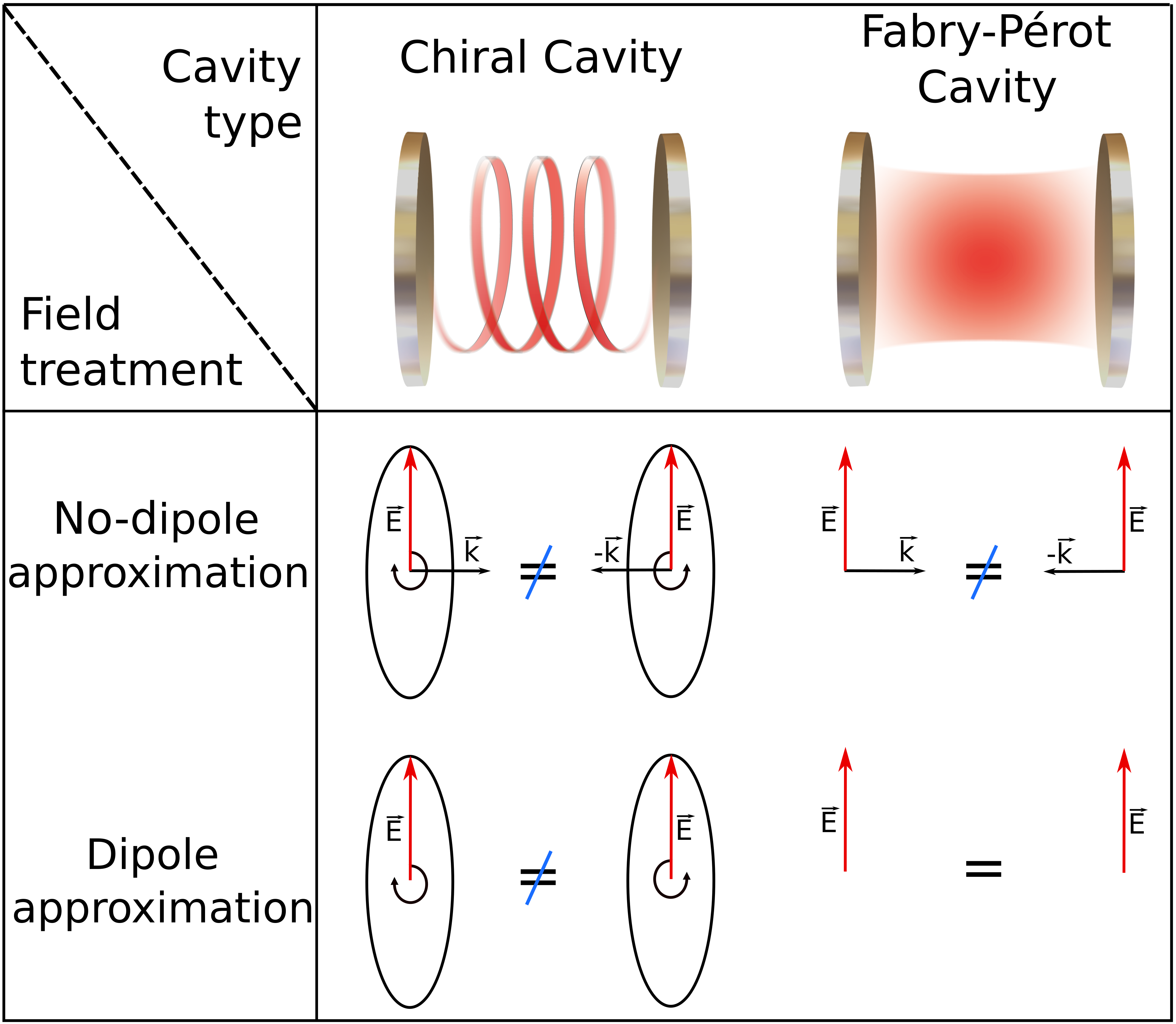}
    \caption{In a chiral cavity (left side of the picture) the left and right moving photons are different even in dipole approximation (the two arrows rotate in a different way) and both the modes must therefore be included in the Hamiltonian. In a linearly polarized cavity, instead, once the dipole approximation has been adopted the contribution from the two photons is exactly the same. Therefore a one mode picture is enough to describe the system.}
    \label{fig:Single}
\end{figure*}
\section{Appendix E: Single mode approximation}
The single-mode approximation is widely used together with the dipole approximation ($e^{\pm i\mathbf{k}\mathbf{r}}=1$)  \cite{rokaj2018light,di2019resolution,schafer2020relevance}. In the single-mode approximation, the Hamiltonian in Eq.~(\ref{eq:Multimode}) becomes
\begin{equation}
\begin{split}
H_{k} =& \frac{1}{2}\sum_{i}\left(\mathbf{p}_{i}-\frac{\lambda}{\sqrt{2\omega_{k}}}\left(\boldsymbol{\epsilon}_{\pm}b_{k}e^{i\mathbf{k}\mathbf{r}_{i}}+\boldsymbol{\epsilon}^{*}_{\pm}b^{\dagger}_{k}e^{-i\mathbf{k}\mathbf{r}_{i}}\right)\right)^{2}\\
+&\sum_{i>j}\frac{1}{\left|r_{i}-r_{j}\right|}+\sum_{I>J}\frac{Z_{I}Z_{J}}{\left|R_{I}-R_{J}\right|}-\sum_{i, I}\frac{Z_{I}}{\left|R_{I}-r_{i}\right|}\\
+&\;\omega_{k}\left(b^{\dagger}_{k}b_{k}+\frac{1}{2}\right),
\label{eq:Singlemode_p}
\end{split}
\end{equation}
or 
\begin{equation}
\begin{split}
H_{k} =& \frac{1}{2}\sum_{i}\left(\mathbf{p}_{i}-\frac{\lambda}{\sqrt{2\omega_{k}}}\left(\boldsymbol{\epsilon}_{\mp}b_{-k}e^{-i\mathbf{k}\mathbf{r}_{i}}+\boldsymbol{\epsilon}^{*}_{\mp}b^{\dagger}_{-k}e^{i\mathbf{k}\mathbf{r}_{i}}\right)\right)^{2}\\
+&\sum_{i>j}\frac{1}{\left|r_{i}-r_{j}\right|}+\sum_{I>J}\frac{Z_{I}Z_{J}}{\left|R_{I}-R_{J}\right|}-\sum_{i, I}\frac{Z_{I}}{\left|R_{I}-r_{i}\right|}\\
+&\;\omega_{k}\left(b^{\dagger}_{-k}b_{-k}+\frac{1}{2}\right),
\label{eq:Singlemode_m}
\end{split}
\end{equation}
depending on which mode is retained, either the one going from left to right or vice versa.
Although these Hamiltonians seem to be theoretically consistent with each other, they do not have the same eigenvalues as there is no unitary transformation connecting them. Choosing one mode of the field only would therefore break the natural symmetry of the Hamiltonian Eq.~(\ref{eq:Multimode}) where left and right are the same.
Therefore, at least two different modes of the electromagnetic field are needed to characterize a chiral cavity.
Specifically, such modes should have the same frequency but opposite wave vector. So far, we have shown that two modes are necessary to describe the chiral field consistently.
A similar line of arguments shows that the two-mode treatment is also required when the dipole approximation is applied in Eq.~(\ref{eq:Multimode}).
A pictorial representation of the results discussed in this section is reported in Fig.~\ref{fig:Single}. In particular, we highlight that, when the field is circularly polarized, the photons moving to the right are different from those moving to the left even in the dipole approximation. Therefore, both modes need to be included explicitly in our description of the vector potential. In the case of a linearly polarized field, instead, the left and right moving photons can only be distinguished if we go beyond the dipole approximation.
\begin{figure*}
    \centering
    \includegraphics[width=\textwidth]{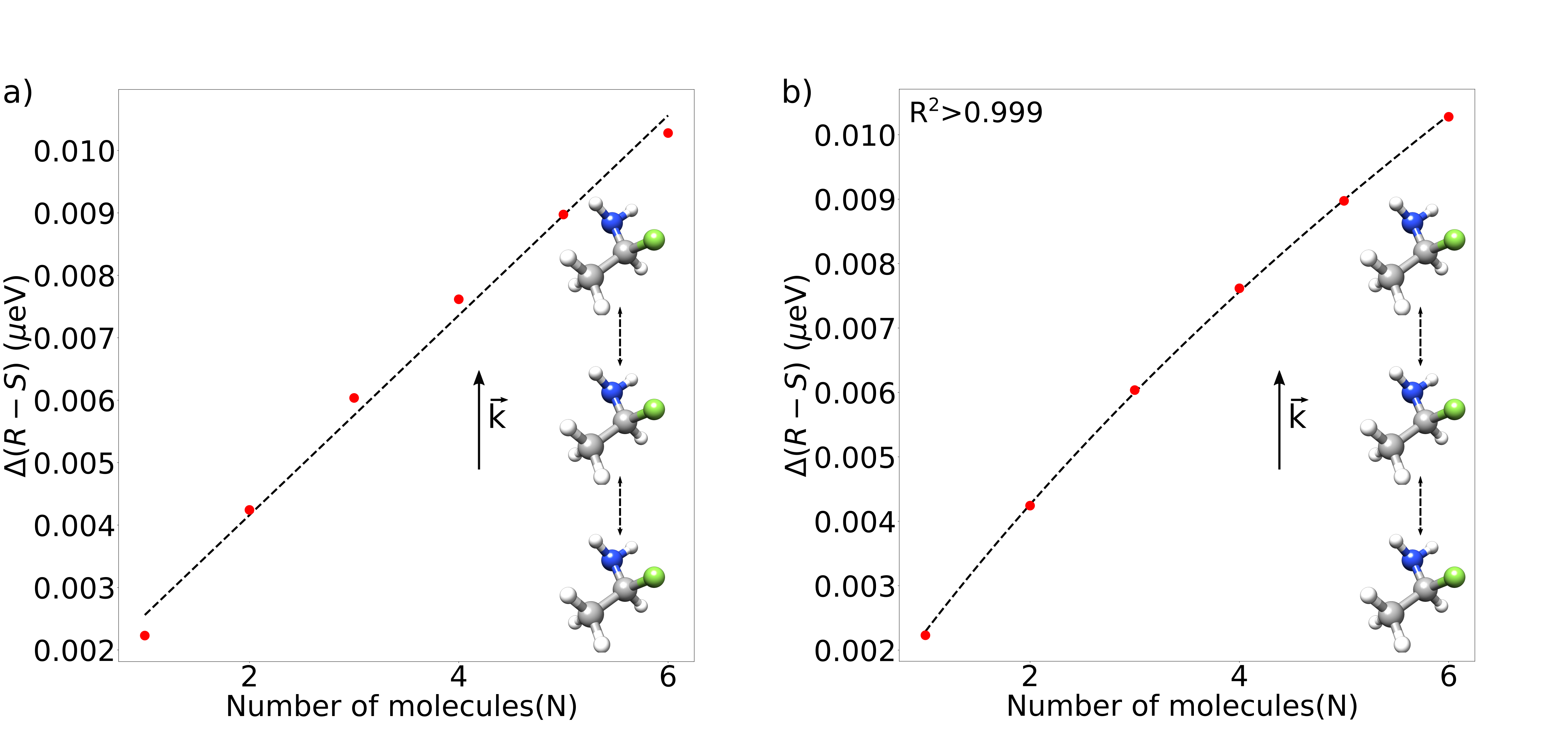}
    \caption{Dispersion of $\Delta \left(R-S\right)$ with respect to the number of chiral centers in the cavity for the case of $\lambda=0.001 \;au$ and $\omega=0.68 \textrm{eV}$. The fitting functions are: a)$\Delta \left(R-S\right)=0.00096 + 0.00160\; N$ and b) $\Delta \left(R-S\right)=-0.00215 + 0.00050 N+ 0.00386 \sqrt{N}$.}
    \label{fig:Christian}
\end{figure*}
\section{Appendix F: Small coupling effects}
In Fig.\ref{fig:Collective}, we discuss how the field induced energy differences are affected by the number of chiral molecules in the cavity. In particular, we found out that the discrimination power grows as the square root of the number of enantiomers in the cavity for a large number of molecules. In this appendix we additionally show that in the limit of small coupling the effects grows almost linearly, see Fig.\ref{fig:Christian}. In this setting, $N_{e}\lambda^{2}<<\omega^{2}$, we have that 
\begin{equation}
    \frac{\lambda}{\sqrt{2\sqrt{\omega^{2}+N_{e}\lambda^{2}}}} \approx \frac{\lambda}{\sqrt{2\omega}}\left(1-\frac{N_{e}\lambda^{2}}{2\omega^{2}}\right).  
\end{equation}
The linearity effect is lost when the number of molecules is such that $N_{e}\lambda^{2}\approx \omega^{2}$ and the $\sqrt{N}$ effect dominates at large number of molecules. This result is in agreement with what is reported in Ref.~\onlinecite{schafer2022chiral}.
\bibliography{bibliography}
\end{document}